\documentclass[11pt, a4paper]{article}
\usepackage{jheppub}
\usepackage{tikz}
\usepackage{bbm}
\usepackage{simplewick}
\usepackage[T1]{fontenc} 
\usepackage{braket}

\newcommand{\be}{\[}
\newcommand{\ee}{\]}
\newcommand{\bq}{\begin{eqnarray}}
\newcommand{\eq}{\end{eqnarray}}

\newcommand{\one}{\hbox{\rm 1\kern-.27em I}}

\title{Intersection of world-lines on curved surfaces and path-ordering of the Wilson loop} 
\author{Chris Curry and Paul Mansfield \\Centre for Particle Theory, University of Durham, Durham DH1 3LE, UK}\emailAdd{p.r.w.mansfield@durham.ac.uk, c.h.curry@durham.ac.uk}
\abstract{
We study contact interactions for long world-lines on a curved surface, focusing on the average number of times two world-lines intersect as a function of their end-points. The result  can be used to extend the concept of path-ordering, as employed in the Wilson loop, from a closed curve into the interior of a surface spanning the curve. Taking this surface as a string world-sheet yields a generalisation of the string contact interaction previously used to represent the Abelian Wilson loop as a tensionless string. We also describe a  supersymmetric generalisation.}


\begin{document}
\maketitle

\section{\bf Introduction}

Maxwell's theory of electromagnetism \cite{Maxwell} takes the fundamental degrees of freedom to be vector fields, however it was inspired in part by Faraday's description \cite{Faraday} in terms of the dynamics of lines of force. The electric field due to a single line of force stretching between a pair of equal charges of opposite sign at $\bf a$ and $\bf b$ can be modelled by Dirac's expression \cite{Dirac}
\be
{\bf E}({\bf x})=\frac{q}{\epsilon_0}\int _C\delta^3 ({\bf x}-{\bf y})\,d{\bf y}\,.\label{1}
\ee
This satisfies Gauss' law, $\epsilon_0\nabla\cdot{\bf E}({\bf x})=q \delta^3 ({\bf x}-{\bf a})-q \delta^3 ({\bf x}-{\bf b})$
but not $\nabla \times {\bf E}=0$ since it is only a part of the electric field of the two charges. Dirac hoped that by dressing electron-positron creation operators by this field 
the divergences of QED could be  softened because when electron-positron pairs are created in the real world they are not created in isolation but are accompanied by electromagnetic fields. Although this single line of force is not the full field of a pair of charges it was  hoped that the full field would result from quantum mechanical averaging. Implementing this idea would be tantamount to taking lines of force to be the degrees of freedom of electromagnetism \cite{Mansfield:2011eq}. Replacing a description in terms of vector fields by one in terms of string-like extended objects is a return to Faraday's point of view. The connection to string theory can be further developed by considering the spacetime generalisation of  (\ref{1}) for the field-strength
\be
F^{\mu\nu}(x)=-q\int_\Sigma \delta^4\left(x-X(\xi)\right)\,
d\Sigma^{\mu\nu}(\xi) \,,\quad
d\Sigma^{\mu\nu}(\xi)=\frac{1}{2}\epsilon^{ab}\partial_a X^\mu
\partial_b X^\nu\,d^2\xi.\label{F}
\ee
$\epsilon^{ab}$ is the antisymmetric tensor density, $X^\mu$ spacetime co-ordinates and $\partial_a$ the derivative with respect to world-sheet co-ordinates $\xi^a$.
$d\Sigma^{\mu\nu}$ is the element of area on the world-sheet $\Sigma$ swept out by the line of force which connects the charges on the boundary curve $B$ associated with the world-line of an electron-positron pair with current density
$$
J^\mu=q\oint_B\delta^4\left(x-w\right) \,dw^\mu\,.
$$
Again Gauss' law $\partial_\mu F^{\mu\nu}=J^\nu$ is satisfied but not  $\epsilon^{\mu\nu\rho\lambda}\partial_\mu F_{\nu\lambda}=0$. Substituting (\ref{F}) into the electromagnetic action $\int d^4x \,F_{\mu\nu}F^{\mu\nu}$ gives 
\be
S^I_4=\kappa\int
d\Sigma^{\mu\nu}(\xi)\,\delta^4\left(X(\xi)-X(\xi')\right)\,
d\Sigma_{\mu\nu}(\xi')
\label{Sem}\nonumber
\ee
The coupling constant $\kappa$ is proportional to the square of the electric charge. 
The argument of the $\delta$-function is non-zero for $\xi=\xi'$ which gives a contribution to $S^I_4$ proportional to the area of $\Sigma$,
i.e. the Nambu-Goto string action, with a divergent coefficient $\propto \delta^4(0)$. When the 
line of force intersects itself the $\delta$-function is again non-zero and so this is a contact interaction. It is Weyl invariant both in the four-dimensional spacetime and on the two-dimensional world-sheet.

(\ref{F}) is not the full field-strength resulting from the charge pair as it represents only a single line of force, but
the full field-strength does emerge after summing over surfaces. In \cite{Edwards:2014xfa}, \cite{Edwards:2014cga} it was shown that the Wilson loop for Abelian gauge theory associated with a closed curve {B} in flat Euclidean space can be written as the partition function of a tensionless four-dimensional string whose world-sheet $\Sigma$ spans {B} with an interaction that is the supersymmetric version of (\ref{Sem}). To see how the contact interaction gives rise to the electromagnetic photon propagator first Fourier decompose the $\delta$-function 
$$
S^I_4=\frac{\kappa}{4}\int_k \,d^2\xi\,d^2\xi'\,
L,\quad \int_k\equiv\int \frac{d^4k}{(2\pi)^4}
$$
\be
L=
\epsilon^{ab}\partial_a X^\mu(\xi)\,
\partial_b X^\nu(\xi)\,e^{ik\cdot \left(X(\xi)-X(\xi')\right)}
\epsilon^{rs}\partial_r X_\mu(\xi')\,\partial_\nu X_s(\xi')\,.
\ee
This has the form of the product of two vertex operators
\[
V^{\mu\nu}_k(\xi)=\epsilon^{ab}\partial_a X^\mu(\xi)\,
\partial_b X^\nu(\xi)\,e^{ik\cdot X(\xi)}\,.
\]
We resolve $\partial X$ into its components along and transverse to $k$ using the projection
$$
{\mathbb{P}}_k(X)^\mu=X^\mu-k^\mu k\cdot X/k^2\,,
$$
so that 
\[
V^{\mu\nu}_k=\epsilon^{ab}\partial_a {\mathbb{P}}_k(X)^\mu\,
\partial_b {\mathbb{P}}_k(X)^\nu\,e^{ik\cdot X}
+2\epsilon^{ab}\partial_a (k\cdot X)k^{[\mu}\,
\partial_b {\mathbb{P}}_k(X)^{\nu ]}\,e^{ik\cdot X}/k^2
\]
\[
=\epsilon^{ab}\partial_a {\mathbb{P}}_k(X)^\mu\,
\partial_b {\mathbb{P}}_k(X)^\nu\,e^{ik\cdot X}
-\partial_a\left(2i\epsilon^{ab} k^{[\mu}\,
\partial_b {\mathbb{P}}_k(X)^{\nu ]}\,e^{ik\cdot X}/k^2\right)
\]
hence
$$
L=\tilde V^{\mu\nu}_k(\xi)\,\tilde V_{\mu\nu\,-k}(\xi')
+2\partial_b \,\partial'_s \left(\epsilon^{ab}\partial_a  {\mathbb{P}}_k(X)^\mu(\xi)\,
 \frac{e^{ik\cdot \left(X(\xi)-X(\xi')\right)}}{k^2}
\epsilon^{rs}\partial_r{\mathbb{P}}_k(X)_\mu(\xi')\right)\, ,
$$
where
$$
\tilde V^{\mu\nu}_k(\xi)\equiv
\epsilon^{ab}\partial_a {\mathbb{P}}_k(X)^\mu\,
\partial_b {\mathbb{P}}_k(X)^\nu\,e^{ik\cdot X}\,.
$$
When substituted back into $S^I_4$ the second term can be written as a double integral over the boundary $B$:
$$
S^I_4=\frac{\kappa}{4}\int_k\,d^2\xi\,d^2\xi'\,
\tilde V_k^{\mu\nu}(\xi)\,\tilde V_{\mu\nu\,\,-k}(\xi')
$$
\be+\frac{\kappa}{2}\int_k
\oint_B \oint_B {\mathbb{P}}_k(dX)^\mu(\xi)\left( \,
\frac{e^{ik\cdot \left(X(\xi)-X(\xi')\right)}}{k^2}\right) 
{\mathbb{P}}_k(dX)_\mu(\xi')\,.\label{resa}
\ee

Averaging over world-sheets (keeping $B$ fixed) using the standard string theory action suppresses the first term leaving the second term which is just the photon propagator integrated over $B$, in other words the expectation value of 
$(\oint dX\cdot A)^2$ in QED.  Suppression of the first term arises because the exponentials $e^{\pm ik\cdot X}$ result in self-contractions so we can write
$$
e^{\pm ik\cdot X}=:e^{\pm ik\cdot X}:\,e^{-\alpha' k^2 G(\xi,\xi)}
$$
where the colons indicate there are no further self-contractions to be made in the enclosed expression when Wick's theorem is applied to evaluate the expectation value. $\alpha'$ is the string scale and $G(\xi,\xi')$ is the Green function for the world-sheet Laplacian. The antisymmetry in $rs$ and the projection operators in $V^{rs}$ prevent any further self-contractions so
$$
V^{\mu\nu}(k,\xi)=:V^{\mu\nu}(k,\xi):\,e^{-\alpha'\pi\, k^2 G(\xi,\xi)}\,.
$$
The Green function at coincident points diverges and should be regulated with a short-distance cut-off, $G(\xi,\xi)\sim -(\log \epsilon)/(2\pi)$ (although it would vanish on the boundary as the curve $B$ is fixed.) If we work in the Wick-rotated theory $k^2>0$ so $e^{-\alpha' \pi \,k^2 G(\xi,\xi)/2}$ is suppressed in the interior of the world-sheet for Fourier modes for which $\alpha' k^2$ is finite as the cut-off is removed. The tensionless limit corresponds to taking $\alpha'/L^2\rightarrow\infty$ where $L$ is a length scale characterising $B$ enhancing the suppression. The suppression is not spoilt by divergences in the expectation value of $V^{rs}_k(\xi)\,V_{rs\,\,-k}(\xi')$ as $\xi$ approaches $\xi'$ \cite{Edwards:2014xfa} as this is just the appearance of the Nambu-Goto action in $\int d^4x \,F_{rs}F^{rs}$. To summarise, if we use $\langle\rangle_{WS}$ to denote the average over world-sheets bounded by $B$, then 
\be
\langle\,S_4^I
\rangle_{WS}=\frac{\kappa}{2}\int_k
\oint_B \oint_B {\mathbb{P}}_k(dX)^\mu(\xi)\left( 
\frac{e^{ik\cdot \left(X(\xi)-X(\xi')\right)}}{k^2}\right) 
{\mathbb{P}}_k(dX)_\mu(\xi')\,.\label{avint}
\ee
The exponential of the right-hand-side of this expression is equal to the expectation value of the Wilson loop, $\exp (iq\oint_B A_r(X) \,dX^r)$ in Abelian gauge theory which is a fundamental object of study, particularly in the first quantised representation of scalar QED \cite{Strassler:1992zr}. 
(The projection ${\mathbb{P}}_k$ appears naturally in Lorenz gauge but can be dropped as the result is gauge invariant).
This suggests that the expectation value of the Wilson loop might be expressed as the world-sheet average of the exponential of $S^I_4$, however divergences appear when the exponential is expanded in powers of $S^I_4$ that potentially spoil the suppression of unwanted terms. In \cite{Edwards:2014xfa} it was shown that these extra terms are absent from the supersymmetric generalisation. Thus the super Wilson loop for (non-supersymmetric) Abelian gauge theory is obtained as the average over world-sheets of the spinning string with a non-standard contact interaction. In first quantisation spin half fermions couple to Abelian gauge fields via this super Wilson loop, so this can be made the starting point for a representation of QED in terms of tensionless 
strings \cite{Edwards:2014cga}. This would provide another approach to first quantised QED which has already been shown to lead to useful results in \cite{Strassler:1992zr}-\cite{Schubert:2001he}.

We would like to generalise the constructions of \cite{Edwards:2014xfa} and \cite{Edwards:2014cga} to curved spacetime, and also to a non-Abelian gauge theory. The problem of the curved background is difficult so in this paper we will study analogous interactions in a lower dimensional model by considering contact interactions for world-lines of particles moving in two dimensions with curved metric $h_{ab}$ (of Euclidean signature). The spacetime Weyl invariant analogue of 
[\ref{Sem}] for a curve $C$ given parametrically by $x^a=x^a(\xi)$ is
\be
\int_{C}
dx^{a}(\xi_1)\,\sqrt {h(x)}\,h_{ab}(x)\,\delta^2\left(x(\xi_1)-x(\xi_2)\right)\,
dx^{b}(\xi_2)\,.
\label{wl}
\ee
This has been investigated in \cite{Edwards:2015hka}.
In two dimensions we can also consider 
\be
n[C]=
\int_{C}
dx^{a}(\xi_1)\,\epsilon_{ab}(x)\,\delta^2\left(x(\xi_1)-x(\xi_2)\right)\,
dx^{b}(\xi_2)
\label{self}
\ee
which counts the number of oriented self-intersections of $C$ or
\be 
n[C_1,\,C_2]=\int_{C_1}\int_{C_2}
{\delta^2(x_1-x_2)}\,\epsilon_{ab}\,{dx_1^a}\,{dx_2^b}
=-n[C_2,\,C_1]\,.\label{nn}
\ee
which counts the number of times two curves intersect.
We will see that this second form of interaction is also of interest in generalising the Wilson loop to a non-Abelian gauge theory, and so this is the interaction we will focus on here. In attempting to represent the Wilson loop in terms world-sheets spanning the loop we would need to find a way of extending the ordering of Lie algebra elements along the boundary into the interior of the world-sheet. We claim that this can be done by counting the number of intersections of random curves, i.e. the expectation value of (\ref{nn}) when we 
average over $C_1$ and $C_2$. Parametrise the world-sheet by the upper half plane with the boundary corresponding to the real axis and consider two curves $C_1$ and $C_2$ in the upper half plane each ending on the boundary,
at $b_1$ and $b_2$ respectively.  We will construct a scale invariant measure for averaging over the curves. Taking into account the skew symmetry of $n[C_1,\,C_2]$ under interchange of $C_1$ and $C_2$ its expectation value will be shown to depend on $b_1$ and $b_2$ as
\be
 \langle n[C_1,\,C_2]\rangle_{C_1,C_2}=k\,(b_1-b_2)/|b_1-b_2|
 \label{boundprop}
\ee
with constant $k$. This function can be used to implement path ordering along the boundary, but also by taking the ends of the curve to move into the interior of the world-sheet we would obtain an extension of path ordering into the body of the world-sheet. Generalising from the Abelian to the non-Abelian theories requires more than just dealing with path ordering as the non-Abelian theory self-interactions make it nonlinear, however these self-interactions are beyond the scope of this paper.  We know that in the Abelian case that supersymmetry is needed to construct the Wilson loop so we will also consider the supersymmmetric generalisation of (\ref{nn}).

\section{Intersection of long world-lines on a surface}

We begin by discussing the purely bosonic model of two curves  on a surface  and study how the average of the number of times they intersect depends on the position of their ends.

Let $C_i$, $i=1,\,2$ be two curves parametrised by $\xi_i$, $0<\xi<1$,  with end-points $a_i$ and $b_i$. The surface $\Sigma$ has intrinsic co-ordinates $x^r$, $r=1,\,2$ metric $h_{rs}(x)$ and anti-symmetric tensor density $\epsilon_{rs}$, so the curves are described by $x^r=x^r_i(\xi_i)$, and their end-points have co-ordinates $a_i^r$ and $b_i^r$. 
For simplicity we take $\Sigma$ to have the topology of a disc and that curves that reach the boundary are reflected specularly.  

We now sum the intersection number (\ref{nn}) over
curves $C_1$ with one end-point, $a_1$, fixed. For any functional $\Omega[C_1]$ consider the functional
integral in which we first integrate over curves with fixed end-points 
$a_1$ and $b_1$ and then integrate over $a_1$
\be
\langle\Omega\rangle_{C_1}\equiv
\int_\Sigma d^2a_1\sqrt {h(a_1)}\,\left\{
\frac{1}{Z}\int {\cal D} {g}\,{\cal D} {x_1}\,
\delta\left(\int_0^1\sqrt{g(\xi)}\,d\xi-T\right)\,\Omega[C_1]\,e^{-S[g,{x_1}]}\right\}\,.\label{avnew2}
\ee
with  \cite{Brink:1976uf}
\be
S[g,{x_1}]=\frac{1}{2}\int_0^1 {g^{-1}(\xi)}\,h_{rs}(x_1)\,\frac{d x^r_1}{d\xi_1}\,\frac{dx_1^s}{d\xi_1}
 \,\sqrt{g(\xi)}\,d\xi \,.\label{bbdvs}
\ee
$g(\xi)$ is a metric-like degree of freedom intrinsic to $C_1$.
To obtain a scale-invariant weight we will take $T\rightarrow \infty$ at the end of our calculation which means we will be looking at curves that are long in terms of the intrinsic metric $g$.

 These expressions are invariant under reparametrisations of $C_1$ so we can choose a gauge in which $g(\xi)$ is constant. The gauge-fixing procedure is the same as in \cite{Mansfield:2011eq} giving
\be
 \langle\,\Omega\,\rangle_{C_1}=\int_\Sigma d^2a_1\sqrt {h(a_1)}\,\left\{\frac{1}{Z}\int {\cal D} {x_1}\,\,\Omega\,e^{-S[{x_1}]}\right\}\,,\label{n}
\ee
with
\be
S[{x_1}]=\frac{1}{2}\int_0^T \,h_{rs}(x_1)\,\dot x^r_1\,\dot x_1^s\,dt\label{action}
\ee
$t=T\xi$ and the dot denotes differentiation with respect to $t$.
We can take some of the $C_2$ dependence of [\ref{n}] outside the functional integral
\[
\langle \,n[C_1,\,C_2]\,\rangle_{C_1}=
\int_{C_2}
\langle \,\int_{C_1}
\delta^2(x_1-x_2)\,d x_1^r 
\,\rangle_{C_1}\,\epsilon_{rs}\,d x_2^s
\]
Introduce a source for the $\delta$-function so that 
\[
\langle \,\int
\delta^2(x_1-x_2)\,\dot x_1^r \,dt
\,\rangle_{C_1}=
\]
\be
\left\{\frac{\delta}{\delta {\cal A}_r({x_2})}\int d^2a_1\sqrt {h(a_1)}\,\frac{1}{Z}
\int {\cal D} {x_1}\,\,e^{-\int_0^T \left(\frac{1}{2}h_{rs}(x_1)\,\dot x^r_1\,\dot x_1^s   -{\cal A}_r\dot x_1^r\right)\,dt
}
\right\}\Big |_{{\bf \cal A}=0}\,.
\label{Eee}
\ee
The functional integral is the path-integral representation of the Euclidean time evolution operator of a particle moving on $\Sigma$ in an electro-magnetic field with vector potential $i{\bf \cal A}$. The classical Hamiltonian is
$\hat H=h^{rs}({ p}+i{ \cal A})_r({p}+i{ \cal A})_s/2$.
For general $h_{rs}$ there is an operator ordering ambiguity in the quantum theory. We resolve this by identifying the 
quantum Hamiltonian with the Laplacian operator acting on scalars because the functional integral is meant to be invariant under general co-ordinate transformations of $x^r$, so the time evolution operator is the heat kernel for this Laplacian on $\Sigma$. We have required that the curves that are summed over are specularly reflected on the boundary which imposes Neumann boundary conditions on the heat-kernel. To see this we will use the method of images. Take the co-ordinates $x^r$ to be points in the upper half-plane and use points in the lower-half-plane to parametrise a surface $\Sigma_R$ attached along the boundary. $\Sigma_R$ is the reflection of $\Sigma$ in the sense that the value of the metric at a point in the lower half-plane is taken to be the value of the metric at the point in the upper half-plane that is its reflection. Any curve $C_1$ from $a_1$ to $b_1$ that is restricted to $\Sigma$ but is reflected once has the 
same Boltzmann factor as a curve that crosses the boundary between $\Sigma$ and $\Sigma_R$ but either starts at $a^R_1$ the reflection of $a_1$ or ends at $b^R_1$ the reflection of $b_1$. Curves that are reflected an even number of times have the same weight as curves from $a_1$ to $b_2$ (or from $a_1^R$ to $b_1^R$) that are not restricted to $\Sigma$ and curves that are reflected an odd number of times have the same weight as curves from $a_1$ to $b_1^R$ (or from $a_1^R$ to $b_1$) that are not restricted to $\Sigma$. So by including reflected curves we are effectively working on the full plane parametrising $\Sigma\cup\Sigma_R$ but including curves with ends that are the reflections of one of the original end-points and so we can identify
\[
 \frac{1}{Z}\int {\cal D} {x_1}\,\,e^{-\int_0^T \left(\frac{1}{ 2}h_{rs}(x_1)\,\dot x^r_1\,\dot x_1^s   -{\cal A}_r\dot x_1^r\right)\,dt
}
=\langle\,b_1\,|\,e^{-T\hat H}\,|\,a_1\,\rangle
+\langle\,b_1\,|\,e^{-T\hat H}\,|\,a_1^R\,\rangle
\]
\be=\langle\,b_1\,|\,e^{-T\hat H}\,|\,a_1\,\rangle
+\langle\,b_1^R\,|\,e^{-T\hat H}\,|\,a_1\,\rangle
={\cal G}_T(b_1,a_1) \,,\label{hatt}
\ee
where 
\[
\langle\,x\,|\, \hat H\,|\,a_1\,\rangle
=-\frac{1}{2\sqrt h}\,(\partial-{\cal A})_r\left({\sqrt h}\,h^{rs}(\partial-{\cal A})_s\,\frac{\delta^2(x-a_1)}{\sqrt{h(a_1)}}
\right)\,.
\]
When either $a_1$ or $b_1$ is on the boundary the derivative of the heat-kernel normal to the boundary vanishes. The bras and kets are normalised to
\[
 \langle\,x\,|\,a_1\,\rangle=\frac{\delta^2(x-a_1)}{\sqrt{h(a_1)}}
\]
and the resolution of the identity involves an integral over 
the whole plane
\[
\int_\Sigma|\,x\,\,\rangle \,\sqrt{h(x)}\,d^2x \,\langle\,x\,|
+\int_{\Sigma_R}|\,x\,\,\rangle \,\sqrt{h(x)}\,d^2x \,\langle\,x\,|
={ 1}\,.
\]
In (\ref{Eee}) we differentiate with respect to the source $\cal A$ at $x_2$, but if we work on the full plane replacing reflected curves restricted to $\Sigma$ by smooth curves on $\Sigma\cup\Sigma_R$ then we should add the derivative with respect to the source $\cal A$ at the reflection of $x_2$, $x_2^R$

Thus (\ref{Eee}) is
\be
\left\{
\frac{\delta}{\delta { \cal A}_r({x_2})}\,\int_\Sigma d^2a_1\sqrt {h(a_1)}\,\,{\cal G}_T(b_1,a_1)\right\}\Big |_{{\bf \cal A}=0}=\label{Eee2}
\ee
\[
-\int_\Sigma d^2a_1\sqrt {h}\,\int_0^T dt\,\langle\,b_1\,|\,e^{-t\hat H_0}
\left(\frac{\delta \hat H}{\delta {\cal A}_r({x_2})}
+\frac{\delta \hat H }{\delta {\cal A}_r({x_2^R})}\right)
\Big |_{{\bf \cal A}=0}e^{(t-T)\hat H_0}\,
\left(\,|\,a_1\,\rangle+\,|\,a_1^R\,\rangle\right).
\]
Where we have set $\hat H_0$ to be $\hat H$ with ${\cal A}=0$.
The resolution of the identity allows us to write
\[
 \frac{\delta \hat H }{\delta {\cal A}_r({x_2})}\Big |_{{\bf \cal A}=0}e^{(t-T)\hat H_0}\,
\left(\,|\,a_1\,\rangle+\,|\,a_1^R\,\rangle\right)
\]
\[
=\int_{\Sigma\cup\Sigma_R}|\,x\,\,\rangle \,\sqrt{h(x)}\,d^2x \,\langle\,x\,|\,\frac{\delta \hat H}{\delta {\cal A}_r({x_2})}\Big |_{{\bf \cal A}=0}e^{(t-T)\hat H_0}\,
\left(\,|\,a_1\,\rangle+\,|\,a_1^R\,\rangle\right)\,.
\]
If we denote $\cal G$ with $\cal A$ set to zero by ${\cal G}^0$
then this becomes
\[
\int_{\Sigma\cup\Sigma_R}|\,x\,\,\rangle \,\frac{d^2x}{2} 
\left(\delta^2(x-x_2)\sqrt h h^{rs}\partial_s\,\,{\cal G}_{T-t}^0(x,a_1)
+\partial_s \left(\sqrt h h^{rs}\delta(x-x_2)\,{\cal G}_{T-t}^0(x,a_1)\right)
\right)
\]
\[
=\frac{1}{2}
 |\,x_2\,\,\rangle \,
\sqrt h\, h^{rs}\partial_s\,\,{\cal G}^0_{T-t}(x_2,a_1)
-\frac{1}{2}\left(\partial_s |\,x_2\,\,\rangle \right)\,\sqrt h\, h^{rs}\,{\cal G}^0_{T-t}(x_2,a_1)\,.
\]
Taking $\Sigma$ to be compact there is a single normalized
zero-mode
\[
 u_0=\frac{1}{\sqrt A},\quad A={\int_\Sigma \sqrt h \,d^2 x},
\]
and since the other eigenfunctions $u_\lambda$ of the Laplacian on $\Sigma$ are orthogonal to $u_0$ it follows that their integrals over $\Sigma$ vanish, so using the spectral decomposition
\[
 \int_\Sigma d^2a_1\,\sqrt{h(a_1)}\,{\cal G}^0_{T-t}(x_2,a_1)
 =\int_\Sigma d^2a_1\,\sqrt{h(a_1)}\,\sum_\lambda
 u_\lambda (x_2) \,e^{-\lambda(T-t)/2}u_\lambda (a_1)=1
\]
so that we can write (\ref{Eee2}) as
\[
\frac{1}{2} \int_0^T dt\,
 \sqrt{h(x_2)}h^{rs}(x_2)\frac{\partial}{\partial x^s_2}\langle\,b_1\,|\,e^{-t\hat H_0}\,\left(|\,x_2\,\rangle
 +|\,x_2^R\,\rangle\right)\,.
\]
The spectral decomposition also fixes the normalisation of (\ref{hatt}) so that $\langle\,1\,\rangle =1$ since
\[
\langle\,1\,\rangle =\int_\Sigma d^2a_1\sqrt {h(a_1)}\,\left\{\frac{1}{Z}\int {\cal D} {x_1}\,e^{-S[{x_1}]}\right\}
=\int_\Sigma d^2a_1\,\sqrt{h(a_1)}\,{\cal G}^0_{T}(x_2,a_1)\,,
\]
(the assumption of compactness is convenient but not actually required here as the same result would follow from conservation of energy applied to the diffusion equation).
Putting this together results in
\be
\langle \,\int
\delta^2(x_1-x_2)\,\dot x_1^r \,dt
\,\rangle_{C_1}=
\frac{1}{2}\,\sqrt {h(x_2)}\,h^{rs}(x_2)\,\frac{\partial}{\partial x_2^r} \,\int_0^T dt\,{\cal G}_t^0(b_1,x_2)\,.
\ee
For $T\rightarrow\infty$ the integral $\int_0^T dt\,{\cal G}_t^0(b_1,x_2)$ is related to the Green function $G$ for the Laplacian with Neumann boundary conditions. 
\[
\int_0^\infty dt\,\left({\cal G}_t^0(b_1,x_2)-\frac{1}{A}\right)=
\sum_{\lambda>0}
 u_\lambda (b_1) \,\frac{1}{\lambda}u_\lambda (x_2)
 =2\,G(b_1,x_2)
 \]
where
\[
 -\frac{1}{\sqrt h}\,\partial_r\left({\sqrt h}\,h^{rs}\partial_s G(x_1,x_2)\right)=
 \frac{\delta^2(x_1-x_2)}{\sqrt h}-\frac{1}{A},
\]
with the Laplacian acting either at $x_1$ or $x_2$.

Keeping $T$ large but finite acts as an infra-red regulator, whilst replacing the lower integration limit by $\epsilon>0$ is a natural way to introduce an ultra-violet regulator. If we now specialise to the conformal gauge so that in complex co-ordinates\footnote{The subscript on the $\delta$-function denotes the use of complex co-ordinates, so if $z=x+iy$ then $\delta^2_c(z)=\frac{1}{2}\delta(x)\,\delta(y)$, $d^2z=2dx\,dy$, $\partial_z(1/\bar z)=2\pi \delta^2_c(z)$.}
\[
 h_{rs}\,dx^r\,dx^s=e^{\phi(z,\bar z)}\,d{\bar z}\,dz, \quad \sqrt h=\frac{ie^{\phi(z,\bar z)}}{2}, \quad -\frac{1}{\sqrt h}\,\partial_r\left({\sqrt h}\,h^{rs}\partial_s f\right)=-4\,e^{-\phi}
 \frac{\partial^2f}{\partial z\partial\bar z}\,,
\]
\[
 n[C_1,C_2]=-i\int_{C1,C2}\left(dz_1d\bar z_2-d\bar z_1dz_2\right)
 \delta_c(z_1-z_2)\,,
\quad
 \int_\Sigma e^{\phi(z,\bar z)}\,d^2z=2A\,,
\]
then 
\[
\int_0^\infty dt\,\left({\cal G}_t^0(z_1,z_2)-\frac{1}{A}\right)
=-\frac{1}{\pi}\log\Big( {| z_1-z_2|}|\bar z_1-z_2|\Big)-\Psi(z_1,z_2)
\]
with $\Psi$ satisfying Neumann boundary conditions and 
\be
-4\,e^{-\phi(z_1)}
 \frac{\partial^2\Psi}{\partial z_1\partial\bar z_1}=-4\,e^{-\phi(z_2)}
 \frac{\partial^2\Psi}{\partial z_2\partial\bar z_2}=\frac{1}{A}\,,
 \label{Psi}
 \ee
and so (for infinite $T$) 
\[
\langle \,\int
\delta^2_c(z_1-z_2)\,d z_1 
\,\rangle_{C_1}=\frac{\partial  G(b_1,z_2)}{\partial\bar z_2}
\]
\be
=
\frac{1}{4\pi}\left(\frac{1}{\bar b_1-\bar z_2}+\frac{1}{ b_1-\bar z_2}\right)-\frac{\partial\Psi}{\partial\bar z_2}\,.\label{avdeltaint}
\ee
Now we can solve (\ref{Psi}) and the boundary conditions to obtain
\be
\frac{\partial\Psi}{\partial\bar z_2}=
\frac{1}{8\pi  A}\int_\Sigma \left(\frac{1}{\bar a-\bar z_2}+
\frac{1}{a-\bar z_2}\right) e^{\phi(a) }\,d^2a\,,
\label{solvpsi}
\ee
so that 
\[
\langle \,n[C_1,\,C_2]\,\rangle_{C_1}=
\]
\be
\frac{i}{4\pi}\int_{C_2}\left(\frac{1}{\bar b_1-\bar z_2}+\frac{1}{ b_1-\bar z_2}\right)
d\bar z_2-\frac{i}{4\pi}\int_{C_2}\left(\frac{1}{b_1-z_2}+\frac{1}{\bar b_1-z_2}\right)d z_2\label{avc1}
\ee
\[
 -i\int_\Sigma\left(\int_{C_2}\left(\frac{1}{\bar a-\bar z_2}+\frac{1}{ a-\bar z_2}\right)
d\bar z_2-\int_{C_2}\left(\frac{1}{a-z_2}+\frac{1}{\bar a-z_2}\right)d z_2\right)\frac{e^{\phi(a) }}{8\pi A}\,d^2a\,.
\]
The integrals over $C_2$ yield logarithms cut along $C_2$, for example
\[
 \frac{i}{4\pi}\int_{C_2}\frac{d\bar z_2}{ \bar b_1-\bar z_2}
d\bar z_2-\frac{i}{4\pi}\int_{C_2}\frac{d z_2}{b_1-z_2}=
-\frac{1}{2\pi}\Im\log_{C_2} \left(\frac {b_1-b_2}{b_1-a_2}\right)\,,
\]

but as we want to average over $C_2$ we can express these as integrals cut along a 
fixed reference curve $C_2^*$ from $a_2$ to $b_2$ plus $2\pi i$ multiples of the winding number about the the points $b_1$ and $a$ of the closed curve made up of $C_2$ and $C_2^*$ reversed.

 \begin{tikzpicture}
 \filldraw[black] (2,2) circle (2pt) node[anchor=west] {$a_2$};
 \filldraw[black] (2,-2) circle (2pt) node[anchor=west] {$b_2$};
 \filldraw[black] (-2.5,-2.2) circle (2pt) node[anchor=east] {$b_1$};
 \filldraw[black] (-2,2.5) circle (2pt) node[anchor=east] {$a$};
 \draw[ultra thick]  (-2.5,-2.2) .. controls (-3.5,0) and (-1
,0) .. (-2,2.5);
\draw[thick] (2,-2) .. controls (-8,-5) and (-4,0.5).. (2,2);
\draw[ultra thick] (2,-2) .. controls (2.7,-0.5) and (0,0.5).. (2,2);
\filldraw[black] (-0.2,-2.2) circle (0pt) node[anchor=west] {$C_2$};
\filldraw[black] (1.7,0) circle (0pt) node[anchor=west] {$C_2^*$};
 \filldraw[black] (-1.7,2) circle (0pt) node[anchor=west] {$C_1^*$};
 \end{tikzpicture}

 \noindent
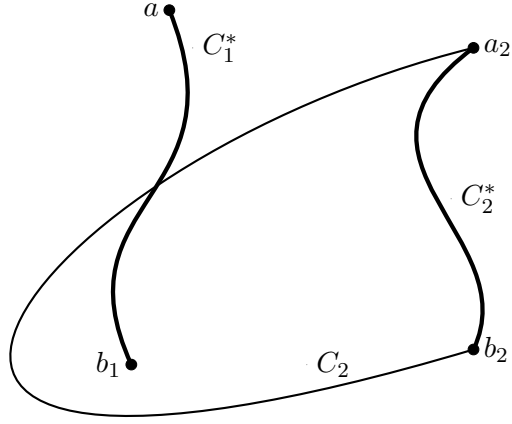
 {{\bf Figure 1} A possible configuration of the curves $C_1^*$, $C_2$ and $C_2^*$ illustrating (\ref{avn})}

\bigskip

The winding numbers can then be written in terms of the number of intersections of  $C_2$ and $C_2^*$ with a reference curve $C_1^*$ from $b_1$ to $a$, so 
\[
\langle \,n[C_1,\,C_2]\,\rangle_{C_1}=
\]
\[
-\frac{1}{2\pi}\Im\log_{C_2^*} \left(\frac {( b_1-b_2)( \bar b_1- b_2)}{(b_1-a_2)(\bar b_1- a_2)}\right)
+\int_\Sigma\Im\log_{C_2^*}\left(\frac {( a-b_2)( \bar a- b_2)}{(a-a_2)(\bar a- a_2)}\right)
\frac{e^{\phi(a) }}{4\pi A}\,d^2a
\]
\be
 -\int_\Sigma
\left(n[C_2,C_1^*]-n[C_2^*,C_1^*]\right)
\frac{e^{\phi(a) }}{2 A}\,d^2a\,,
\label{avn}
\ee
where the subscript denotes that the logarithms, viewed as functions of $b_1$, $a$ and their complex conjugates  are cut along $C_2^*$. 
The only dependence on $C_2$ is via $n[C_2,C_1^*]$ and $a_2$ so if we now average over $C_2$ and its end-point $a_2$ using
\[
\langle \,n[C_2,\,C_1^*]\,\rangle_{C_2}=
\]
\[
-\frac{1}{2\pi}\Im\log_{C_1^*} \left(\frac {( b_2-b_1)( \bar b_2- b_1)}{(b_2-a)(\bar b_2- a)}\right)
 +\int_\Sigma\Im\log_{C_1^*}\left(\frac {( a_2-b_1)( \bar a_2- b_1)}{(a_2-a)(\bar a_2- a)}\right)
\frac{e^{\phi(a_2) }}{4\pi A}\,d^2a
\]
where now the subscript denotes that the logarithms, viewed as functions of $b_2$, $a_2$ and their complex conjugates  are cut along $C_1^*$. 

Observe that the following difference in logarithms cut along $C_1^*$ and $C_2^*$ is proportional to the number of times 
$C_1^*$ and $C_2^*$ intersect:
\be
\log_{C_2^*} \left(\frac {( b_1-b_2)
(a-a_2)
}{(b_1-a_2)( a-b_2)
}\right)
-\log_{C_1^*} \left(\frac {( b_2-b_1)
(a_2-a)
}{(a_2-b_1)( b_2-a)
}\right)
=2\pi i
n[C_2^*,C_1^*]\,.\label{contour}
\ee
This is illusrated in Figure 2.

\begin{tikzpicture}
 \filldraw[black] (2,2) circle (2pt) node[anchor=west] {$a_2$};
 \filldraw[black] (1.85,1.85) circle (0pt) node[anchor=east] {$\alpha_2$};
 \filldraw[black] (2,-2) circle (2pt) node[anchor=west] {$b_2$};
 \filldraw[black] (1.85,-1.85) circle (0pt) node[anchor=east] {$\beta_2$};
 \filldraw[black] (-2.5,-2.2) circle (2pt) node[anchor=east] {$b_1$};
 \filldraw[black] (-2.25,-2) circle (0pt) node[anchor=west] {$\beta_1$};
 \filldraw[black] (-2,2.5) circle (2pt) node[anchor=east] {$a$};
 \filldraw[black] (-1.8,2.25) circle (0pt) node[anchor=west]{$\bf \alpha_1$};\filldraw[black] (-0.15,0.45) circle (0pt) node[anchor=west]{$\bf \theta$};
 \draw (2,2) -- (-2.5,-2.2);
 \draw (2,2) -- (-2,2.5); 
\draw (2,-2) -- (-2,2.5); 
\draw (2,-2) -- (-2.5,-2.2);
\draw[ultra thick]  (2,-2) .. controls (2.7,-0.5) and (0.7,0.5).. (2,2);
\draw[ultra thick]  (-2.5,-2.2) .. controls (-3.5,0) and (-1
,0) .. (-2,2.5);
 
 \filldraw[black] (9,2) circle (2pt) node[anchor=west] {$a_2$};
 \filldraw[black] (8.85,1.85) circle (0pt) node[anchor=east] {$\alpha_2$};
 \filldraw[black] (9,-2) circle (2pt) node[anchor=west] {$b_2$};
 \filldraw[black] (8.85,-1.85) circle (0pt) node[anchor=east] {$\beta_2$};
 \filldraw[black] (4.5,-2.2) circle (2pt) node[anchor=east] {$b_1$};
 \filldraw[black] (4.75,-2) circle (0pt) node[anchor=west] {$\beta_1$};
 \filldraw[black] (5,2.5) circle (2pt) node[anchor=east] {$a$};
 \filldraw[black] (5.2,2.25) circle (0pt) node[anchor=west]{$\bf \alpha_1$};
 \draw (9,2) -- (4.5,-2.2);
 \draw (9,2) -- (5,2.5); 
\draw (9,-2) -- (5,2.5); 
\draw (9,-2) -- (4.5,-2.2);
\draw[ultra thick] (9,-2) .. controls (9.7,-0.5) and (7.7,0.5).. (9,2);
\draw[ultra thick] (4.5,-2.2) .. controls (12,1.5) and (11
,4) .. (5,2.5);

\end{tikzpicture}

\noindent
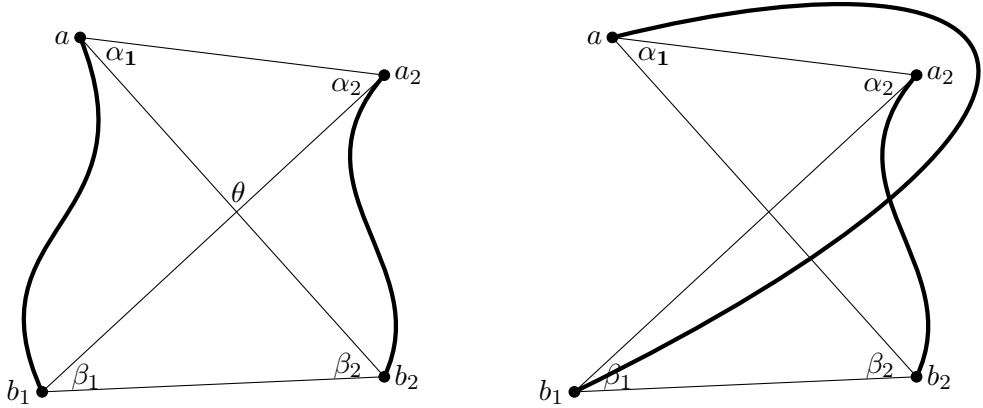
{{\bf Figure 2} Two configurations of the curves $C_1^*$ and $C_2^*$.

\bigskip

The angle swept out by the line from $z$ to $b_1$ as $z$ moves along $C_2^*$ from $a_2$ to $b_2$ is the imaginary part of 
$
\log_{C_2^*} \left({( b_1-b_2)/
}{(b_1-a_2)
}\right)
$
which is $-\beta_1$ in both figures. Similarly the angle swept out 
by the line from $z$ to $a$ is the imaginary part of
$
\log_{C_2^*} \left({( a-b_2)/
}{(a-a_2)
}\right)
$
which is $-\alpha_1$ in both figures. The imaginary part of
$
 \log_{C_1^*} \left({( b_2-b_1)
(a_2-a)
}/{(( b_2-a)(a_2-b_1))
}\right)
$
is the difference in the angles swept out by the lines from $z$ to $b_2$ and from $z$ to $a_2$ as $z$ moves along $C_1^*$ from $a$ to $b_1$. For the left hand figure, in which the curves $C_1^*$ and $C_2^*$ do not intersect, this is $\beta_2-\alpha_2$.
In the right hand figure the line from $z$ to $a_2$ sweeps out $-(2\pi-\alpha_2)$ so the difference in the two angles is $\beta_2+
(2\pi-\alpha_2)$. Also, in the right hand figure the curves $C_1^*$ and $C_2^*$  intersect with $n[C_2^*,C_1^*]=-1$, so for the two figures (\ref{contour}) is
\[
 \alpha_1-\beta_1-(\beta_2-\alpha_2)=0, \quad{\rm and}\quad
 \alpha_1-\beta_1-(\beta_2+2\pi-\alpha_2)=-2\pi\,,
\]
both of which hold because $\alpha_1$, $\alpha_2$ and $\theta$ are the angles of the top triangle in the figure and $\beta_1$, $\beta_2$ and $\theta$ are the angles in the lower triangle.

Using (\ref{contour}) we are just left with 
\[
\langle \,\langle \,n[C_1,\,C_2]\,\rangle_{C_1}\,\rangle_{C_2}=
\]
\be
-\Im\int_\Sigma\left(
 \log_{C_2^*}\frac{(\bar b_1-b_2)(\bar a-a_2)}{(\bar a -b_2)(\bar b_1-a_2)}-\log_{C_1^*}\frac{(\bar b_2-b_1)(\bar a_2-a)}{(\bar b_2-a)(\bar a_2-b_1)} \right)
 \frac{e^{\phi(a)+\phi(a_2) }}{8\pi A^2}\,d^2a\,d^2a_2\,.\label{avav}
\ee
We can now interpret this expression in the light of the comments relating to path-ordering along the boundary using (\ref{boundprop}). Let $b_1$ and $b_2$ approach the real axis so
$b_1=x_1+i\epsilon_1$ and $b_2=x_2+i\epsilon_2$ with $x_1$ and $x_2$ real, and denote by $G(x_1,x_2)$ the resulting value of $\langle \,\langle \,n[C_1,\,C_2]\,\rangle_{C_1}\,\rangle_{C_2}$
then 
\[
 \log_{C_2^*}(\bar b_1-b_2)-\log_{C_1^*}(\bar b_2-b_1)=i\pi \frac {x_1-x_2}{|x_1-x_2|}\,.
\]
As this is independent of $a$ and $a_2$ the area integrals in (\ref{avav}) can be done to give 
\[
G(x_1,x_2)=
 -\frac {x_1-x_2}{2|x_1-x_2|}+F(x_1)-F(x_2)\,,
\]
which is (\ref{boundprop}) apart from the function $F$. To interpret $F$ differentiate with respect to $x_1$
\be
\frac{\partial}{\partial x_1}G(x_1,x_2)
=
-\delta(x_1-x_2)+F'(x_1)\,.\label{gf1}
\ee
The real axis parametrises the boundary of $\Sigma$ which has finite length and the co-ordinates $x=\pm\infty$ describe the same point on this boundary so
for consistency we should have
\be
 0=\int_{-\infty}^{\infty}\frac{\partial}{\partial x_1}
\langle \,\langle \,n[C_1,\,C_2]\,\rangle_{C_1}\,\rangle_{C_2}\,dx_1=
-1+\int_{-\infty}^{\infty}F'(x_1)\,dx_1\,,
\ee
but from (\ref{avav})
\[
 F'(x_1)=\Im\int_\Sigma\left(\frac{1}{x_1-a_2}
 -\frac{1}{x_1-\bar a_2}
 \right)
 \frac{e^{\phi(a_2) }}{4\pi A}\,d^2a_2
\]
which does indeed integrate to $+1$. Now (\ref{gf1}) is a Green function equation for $\partial/\partial x$ on a closed loop, i.e. the propagator for a one-dimensional field $\psi$ with action $\int dx \tilde\psi \psi'$. This field theory has been used to represent path-ordering around the loop in \cite{Samuel:1978iy}-\cite{Bastianelli:2013pta}. Since $G(x_1,x_2)$ is just the boundary value of the average of the intersection number we have a natural way of extending path ordering into the interior of $\Sigma$. This extension coincides with the propagator of the topological field theory constructed in \cite{Broda:1995wv} for just this purpose. To see this connection note that in Broda's model the boundary field $\psi$ is assumed to be the boundary value of a bulk field and the extension into the bulk can be done arbitrarily giving rise to a topological field theory with invariance $\delta\psi=\theta$ with $\theta$ being any function vanishing on the boundary. This invariance is gauge-fixed by requiring $\psi$ 
to be harmonic.  Just as in the topological theory the average intersection number satisfies Laplace's equation in the bulk because it is non-singular as $b_1$ approaches $b_2$ in the interior.
So taking 
$\langle \,\langle \,n[C_1,\,C_2]\,\rangle_{C_1}\,\rangle_{C_2}$ as the propagator for new variables in the interior of $\Sigma$ 
might provide a way of building Lie algebraic structure into the contact interaction $S^I_4$. However such additional degrees of freedom need to generate the three and four point self-interactions of Yang-Mills theory as well as just producing path-ordering of Lie algebra elements. These interactions are additional to those of Abelian gauge theories and would arise from extra divergences when the vertices in $S^I_4$ approach each other on $\Sigma$. Their study is beyond the scope of this paper.

\section{Path-ordering in a string representation of the Wilson loop.}

Ultimately we would like to generalise the result of \cite{Edwards:2014xfa} and \cite{Edwards:2014cga} and construct a representation of the expectation value of the Wilson loop for Yang-Mills theory in terms of tensionless strings with contact interactions. We will not be able to do this here as we know that we would have to account for the self-interactions of the Yang-Mills field. However we will be able to see the seeds of some of the extra structures needed in the non-Abelian theory appearing in the bosonic theories we have considered.

If we ignore the self-interactions of Yang-Mills theory then the expectation value of the Wilson loop $\langle {\cal P}\,\exp (-q\oint_B \tau^JA_\mu^J(X) \,dX^\mu)\rangle$ for a non-Abelian gauge theory with (anti-Hermitian) Lie algebra generators $\tau^J$ is, in Lorenz gauge,
\[
{\rm Tr} \,{\cal P}\,\exp\left(
\frac{\kappa}{ 2}\int_k
\oint_B \oint_B \tau^J\,{\mathbb{P}}_k(dX)^\mu(\xi)
\left( 
\frac{e^{ik\cdot \left(X(\xi)-X(\xi')\right)}}{k^2}\right) \tau^J\,{\mathbb{P}}_k(dX)_\mu(\xi')\right)
\]
which differs from the result in the Abelian theory just by the path-ordering of the Lie algebra generators. This path-ordering can be replaced by a functional integral over an anti-commuting field that on $B$ \cite{Samuel:1978iy}
\[
\int {\cal D}(\psi^\dagger,\psi)\,\psi^\dagger(1)\,\psi(0)\,\exp\Bigg(\int_0^1 \psi^\dagger \dot\psi\,dt+
\]
\be
\frac{\kappa}{ 2}\int_k
\oint_B \oint_B \left(\psi^\dagger\tau^J\psi\,{\mathbb{P}}_k(dX)^\mu\right)|_\xi
\left( 
\frac{e^{ik\cdot \left(X(\xi)-X(\xi')\right)}}{k^2}\right) \left(\psi^\dagger\tau^J\psi\,{\mathbb{P}}_k(dX)_\mu\right)|_{\xi'}\Bigg)\,.
\label{ssam}
\ee
Apart from the kinetic term for $\psi$ this differs from the  Abelian case (\ref{avint}) by the inclusion of the Lie algebra terms
$J^A\equiv \psi^\dagger\tau^A\psi$. This
suggests that the non-Abelian generalisation of the contact interaction (\ref{Sem}) should also be modified to include the 
$J^A$ and take the form 
\[
S^{YM}_4=\kappa\int\left(J^A\,
d\Sigma^{\mu\nu}\right)|_\xi\,\delta^4\left(X(\xi)-X(\xi')\right)\,
\left(J^A\,
d\Sigma_{\mu\nu}\right)|_{\xi'}
\]
\be
=\frac{\kappa}{4}\int_k\int_\Sigma\int_\Sigma\left(J^A\,V_k^{\mu\nu}\right)|_{\xi_1}
\left(J^A\,V_{\mu\nu\,-k}\right)|_{\xi_2}\,d^2\xi_1\,d^2\xi_2
\label{Semnona}
\ee
where we now extend the meaning of $\psi^\dagger$ and $\psi$ from anti-commuting boundary fields to anti-commuting variables on the world-sheet with propagator given by the average of the intersection number:
\[
\contraction{}{\psi_R^\dagger }{(b_1)\,}{\psi_S}\psi_R ^\dagger(b_1)\,\psi_S (b_2)=\langle\, \langle\, n[C_1,C_2]\,\rangle_{C_2'}\,\rangle_{C_1}\,\delta_{RS}
\]
because this reduces to the propagator for (\ref{ssam}) when $b_1$ and $b_2$ are on the boundary. These extra terms in the contact interaction modify (\ref{resa}). As in Section One we use the projector ${\mathbb{P}}_k$ to write 
\[
J^A\,V^{\mu\nu}_k
=J^A\,\epsilon^{ab}\partial_a {\mathbb{P}}_k(X)^\mu\,
\partial_b {\mathbb{P}}_k(X)^\nu\,e^{ik\cdot X}+
\]
\be
(\partial_aJ^A)\,\left(2i\epsilon^{ab} k^{[\mu}\,
\partial_b {\mathbb{P}}_k(X)^{\nu ]}\,e^{ik\cdot X}/k^2\right)
-\partial_a\left(2iJ^A\,\epsilon^{ab} k^{[\mu}\,
\partial_b {\mathbb{P}}_k(X)^{\nu ]}\,e^{ik\cdot X}/k^2\right)\,,
\label{witho}
\ee
so that the contact interaction becomes
$$
S^{YM}_4=\frac{\kappa}{ 4}\int_k\,\Bigg(\int_\Sigma \int_\Sigma d^2\xi\,d^2\xi'\,
\left(J^A\,\tilde V_k^{\mu\nu}\right)|_\xi\,\left(J^A \,\tilde V_{\mu\nu\,\,-k}\right)|_{\xi'}
$$
\[
+{2}\int_\Sigma \int_\Sigma d^2\xi\,d^2\xi'\,
  \left(\partial_a  J^A\,  \epsilon^{ab}\partial_b  {\mathbb{P}}_k(X)^\mu e^{ik\cdot X}  \right)|_\xi\,
\,\frac{1}{k^2} \,\left(\partial_r J^A\,
\epsilon^{rs}\partial_s{\mathbb{P}}_k(X)_\mu e^{-ik\cdot X}  \right)|_{\xi'}
\]
\[
+{4}\int_\Sigma\oint_B d^2\xi\,
  \left(\partial_a  J^A\,  \epsilon^{ab}\partial_b  X^\mu e^{ik\cdot X}
  \right)|_\xi\,
 \,\frac{1}{k^2}\, \left(J^A\,
{\mathbb{P}}_k(dX)_\mu e^{-ik\cdot X} \right)|_{\xi'}
\]
\be+
2\oint_B \oint_B \left(J^A\,{\mathbb{P}}_k(dX)^\mu e^{ik\cdot X} \right)|_\xi\,
\frac{1}{k^2}\, 
\left(J^A\,   {\mathbb{P}}_k(dX)_\mu  e^{-ik\cdot X}\right)_{\xi'}\Bigg)\,.\label{resa'}
\ee
The last term depends only on the boundary values of $X$ so is unchanged if we average over world-sheets spanning $B$.  The other terms will be suppressed due to self-contractions in the exponential just as in the Abelian case, so
\be
\langle\,S_4^{YM}
\rangle_{WS}=
\frac{\kappa}{ 2}\int_k
\oint_B \oint_B \left(J^A\,{\mathbb{P}}_k(dX)^\mu\right)|_\xi \left( 
\frac{e^{ik\cdot \left(X(\xi)-X(\xi')\right)}}{k^2}\right) 
\left(J^A\,   {\mathbb{P}}_k(dX)_\mu\right)|_{\xi'}\,.\label{freenon}
\ee
If this were to exponentiate then we would obtain the exponential in (\ref{ssam})), however we know from the Abelian case that world-sheet supersymmetry is required to eliminate extra divergences when there are products of interactions \cite{Edwards:2014xfa}-\cite{Edwards:2014cga}. This supersymmetry should extend to the $\psi$ degrees of freedom so in Section 5 we seek a supersymmetric formulation of the $\psi$.

\section{Bosonic generalisations}

Having computed the average intersection number for curves with one end fixed we can readily modify the calculation to calculate the average intersection number for curves with both ends fixed and also the average self-intersection number of a single curve as well as the two-dimensional contact interaction mentioned in the introduction. We consider all of these in this section.

\subsection{Intersection number of two curves with both ends fixed}

If we do not integrate over the ends of the curves $a_i$ then (\ref{Eee}) becomes
\be
\langle \,\int
\delta^2(x_1-x_2)\,\dot x_1^u \,dt
\,\rangle_{C_1}=
\left\{\frac{\delta}{\delta {\cal A}_u ({x_2})}\frac{1}{Z'}
\int {\cal D} {x_1}\,\,e^{-\int_0^T \left(\frac{1}{ 2}h_{rs}(x_1)\,\dot x^r_1\,\dot x_1^s   -{\cal A}_r\dot x_1^r\right)\,dt
}
\right\}\Big |_{{\bf \cal A}=0}\,.
\label{Eee'}
\ee
with the normalisation constant changing its value so that we still have $\langle \,\int
1\,\rangle_{C_1}=1$ despite changing our averaging process. (\ref{Eee2}) is replaced by

\be
\frac{\delta {\cal G}_T(b_1,a_1)}{\delta { \cal A}_r({x_2})}\Big |_{{\bf \cal A}=0}
\ee
\[
=
-\frac{1}{Z'}\,\int_0^T dt\,\langle\,b_1\,|\,e^{-t\hat H_0}
\left(\frac{\delta \hat H }{\delta {\cal A}_r({x_2})}
+\frac{\delta \hat H }{\delta {\cal A}_r({x_2^R})}\right)
\Big |_{{\bf \cal A}=0}e^{(t-T)\hat H_0}\,
\left(\,|\,a_1\,\rangle+\,|\,a_1^R\,\rangle\right)
\]
\[
 =\frac{1}{2Z'}\int_0^T dt\,\left({\cal G}^0_t(b_1,x_2)\sqrt h h^{rs}\partial_s {\cal G}^0_{T-t}(x_2,a_1)-\partial_s{\cal G}^0_t(b_1,x_2)\sqrt h h^{rs} {\cal G}^0_{T-t}(x_2,a_1)
 \right)\,.
\]
Since we are taking the limit of $T\rightarrow\infty$ the integration over $t$ can be split into two pieces, one where $t$ is close to $0$ and the other where $t$ is close to $T$. For the first region the spectral decomposition allows us to set ${\cal G}^0_{T-t}(x_2,a_1)=1/A$ and for the second region we can set ${\cal G}^0_t(b_1,x_2)=1/A$. We can then extend the integration regions to obtain 
\newpage
\[
\lim_{T\rightarrow \infty} \left\{
\frac{\delta}{\delta { \cal A}_r({x_2})}\,\int_\Sigma da_1^1da_1^2\sqrt {h(a_1)}\,\,{\cal G}_T(b_1,a_1)\right\}\Big |_{{\bf \cal A}=0}
\]
\[
 =\frac{1}{2AZ'}\int_0^\infty dt\,\left(-\sqrt h h^{rs} \partial_s{\cal G}^0_t(b_1,x_2)+\sqrt h h^{rs}\partial_s {\cal G}^0_{T-t}(x_2,a_1)
 \right)
\]
in which we see the appearance of the Green function again. The spectral decomposition fixes the value of $Z'$ as 
\[
\langle\,1\,\rangle =\lim_{T\rightarrow\infty}\,\left\{\frac{1}{ Z'}\int {\cal D} {x_1}\,e^{-S[{x_1}]}\right\}
=\lim_{T\rightarrow\infty}\,\frac{1}{ Z'}{\cal G}^0_{T}(x_2,a_1)\,=\frac{1}{Z'A},
\]
so we get as the generalisation of (\ref{avdeltaint})
\[
\langle \,\int
\delta^2_c(z_1-z_2)\,d z_1 
\,\rangle_{C_1}=\frac{\partial  G(b_1,z_2)}{\partial\bar z_2}-\frac{\partial  G(z_2,a_1)}{\partial\bar z_2}
\]
\be
=
\frac{1}{4\pi}\left(\frac{1}{\bar b_1-\bar z_2}+\frac{1}{ b_1-\bar z_2}\right)
-\frac{1}{4\pi}\left(\frac{1}{\bar a_1-\bar z_2}+\frac{1}{ a_1-\bar z_2}\right)\label{ba}
\ee
which gives 
\[
\langle \,n[C_1,\,C_2]\,\rangle_{C_1}=
\]
\[
-\frac{1}{2\pi}\Im\log_{C_2^*} \left(\frac {( b_1-b_2)( \bar b_1- b_2)}{(b_1-a_2)(\bar b_1- a_2)}\right)
+\frac{1}{2\pi}\Im\log_{C_2^*}\left(\frac {( a_1-b_2)( \bar a_1- b_2)}{(a_1-a_2)(\bar a_1- a_2)}\right)
\]
\[
-\left(n[C_2,C_1^*]-n[C_2^*,C_1^*]\right)
 \]
so on averaging over $C_2$ keeping its ends fixed we finally arrive at
\[
\langle \,\langle \,n[C_1,\,C_2]\,\rangle_{C_1}\,\rangle_{C_2}=
\]
\be-\frac{1}{2\pi}
\Im\left(
 \log_{C_2^*}\frac{(\bar b_1-b_2)(\bar a_1-a_2)}{(\bar a_1 -b_2)(\bar b_1-a_2)}-\log_{C_1^*}\frac{(\bar b_2-b_1)(\bar a_2-a_1)}{(\bar b_2-a_1)(\bar a_2-b_1)} \right)\,.
 \label{fep}
\ee
Averaging this over the points $a_1$, $a_2$ takes us back to (\ref{avav}).

Another representation of the average intersection number with fixed end-points is obtained by starting from
\[
\langle \,\langle \,n[C_1,\,C_2]\,\rangle_{C_1}\,\rangle_{C_2}=-i\,
\int_M
d^2z\,\Big\{\langle \,\int
\delta^2_c(z_1-z)\,d z_1 
\,\rangle_{C_1}\,\langle \,\int
\delta^2_c(z_2-z)\,d \bar z_2 
\,\rangle_{C_2}
\]
\[
-
\langle \,\int
\delta^2_c(z_1-z)\,d \bar z_1 
\,\rangle_{C_1}\,\langle \,\int
\delta^2_c(z_2-z)\,d  z_2 
\,\rangle_{C_2}\,,
\Big\}
\]
where $M$ is the upper half plane. 
Using (\ref{ba}) this becomes 
\[
H\equiv
 i\int_M \frac{d^2z}{16\pi^2}\Big\{\left(
 \frac{1}{\bar b_1-\bar z}+\frac{1}{b_1-\bar z}
 -\frac{1}{\bar a_1-\bar z}-\frac{1}{a_1-\bar z}
 \right)\times
 \]
 \[\left(
 \frac{1}{b_2- z}+\frac{1}{\bar b_2-z}
 -\frac{1}{a_2-z}-\frac{1}{\bar a_2-z}
 \right)
\]
\[
 -\left(
 \frac{1}{b_1-z}+\frac{1}{\bar b_1-z}
 -\frac{1}{a_1-z}-\frac{1}{\bar a_1-z}
 \right)\left(
 \frac{1}{\bar b_2- \bar z}+\frac{1}{b_2-\bar z}
 -\frac{1}{\bar a_2-\bar z}-\frac{1}{a_2-\bar z}
 \right)\Big\}\,.
\]
We can check that this agrees with (\ref{fep}) by differentiating with respect to $b_1$ 
\[
 \frac{\partial H }{\partial b_1}=
\]
\[
-i \int_M \frac{d^2z}{16\pi^2}\,\Bigg\{
\left(-2\pi \delta_c(b_1-z)-
\frac{\partial  }{\partial  b_1 }
 \frac{1}{b_1-\bar z}\right)
 \left(
 \frac{1}{b_2- z}+\frac{1}{\bar b_2-z}
 -\frac{1}{a_2-z}-\frac{1}{\bar a_2-z}
 \right)
\]
\[
+ \left(\frac{\partial  }{\partial b_1}\frac{1}{b_1-z}\right)\left(
 \frac{1}{\bar b_2- \bar z}+\frac{1}{b_2-\bar z}
 -\frac{1}{\bar a_2-\bar z}-\frac{1}{a_2-\bar z}
 \right)\Bigg\}
 \]
\[
=-i\int_M \frac{d^2z}{16\pi^2}\,\Bigg\{
\frac{\partial  }{\partial  \bar z }\left(
 \frac{1}{b_1-\bar z}
 \left(
 \frac{1}{b_2- z}+\frac{1}{\bar b_2-z}
 -\frac{1}{a_2-z}-\frac{1}{\bar a_2-z}
 \right)\right)
\]
\[+2\pi \left(\delta_c(b_2-z)-\delta_c(a_2-z)\right)\frac{1}{b_1-\bar z}
\]
\[
- \frac{\partial  }{\partial z}\left(\frac{1}{b_1-z}\left(
 \frac{1}{\bar b_2- \bar z}+\frac{1}{b_2-\bar z}
 -\frac{1}{\bar a_2-\bar z}-\frac{1}{a_2-\bar z}
 \right)\right)
\]
\[
-2\pi \left(\delta_c(b_2-z)-\delta_c(a_2-z)\right)\frac{1}{b_1-z}
\Bigg\}
 \]
\[
+\frac{i}{8\pi} \left(
 \frac{1}{b_2- b_1}+\frac{1}{\bar b_2-b_1}
 -\frac{1}{a_2-b_1}-\frac{1}{\bar a_2-b_1}
 \right)\,.
\]
Using Stokes' theorem this becomes
\[
\frac{i}{4\pi} \left(
\frac{1}{\bar b_2-b_1}
 -\frac{1}{\bar a_2-b_1}
 \right)
  -\frac{1}{8\pi^2}\int_{-\infty}^\infty \frac{dx}{b_1-x}  \left(\frac{1}{b_2-x}+\frac{1}{\bar b_2-x}-\frac{1}{a_2-x}-\frac{1}{\bar a_2-x}\right)
   \]
and finally computing the integral by closing the contour above the real axis results in 
 \[
 \frac{\partial H }{\partial b_1}=
\frac{i}{2\pi} \left(
\frac{1}{\bar b_2-b_1}
 -\frac{1}{\bar a_2-b_1}
 \right)\,,
\]
which coincides with the derivative of (\ref{fep}).

 \subsection{Self-intersections of a single curve}
 The average self-intersection number for a curve $n[C]$ (\ref{self}) with fixed ends $a$ and $b$ can be represented in a similar way to that of $n[C_1,C_2]$ in the previous subsection
\[
\langle \,n[C]\,\rangle_{C}=
\int_M
d^2y\int\langle \,
\delta^2(x(\xi_1)-y)\,\epsilon_{rs} \dot x^r(\xi_1)\dot x^s(\xi_2)
\delta^2(x(\xi_2)-y)
\,\rangle_{C}\,d\xi_1\,d\xi_2\,.
\]
We now have to deal with two insertions, but as before they can be obtained by functional differentiation with respect to a source
\newpage
\[
\langle \,n[C]\,\rangle_{C}=
\]
\[\lim_{T\rightarrow\infty}\int_M
d^2y
\left\{\epsilon_{rs}\frac{\delta}{\delta {\cal A}_r({y})}\frac{\delta}{\delta {\cal A}_s({y})}\frac{1}{Z}
\int {\cal D} {x_1}\,\,e^{-\int_0^T \left(\frac{1}{2}h_{pq}(x_1)\,\dot x^p_1\,\dot x_1^q   -{\cal A}_p\dot x_1^p\right)\,dt
}
\right\}\Big |_{{\bf \cal A}=0}
\]
\[
 =\lim_{T\rightarrow\infty}\frac{1}{4Z}\int_M
d^2y\,\int_0^T dt_1\,\int_\epsilon^T dt_2\,\times
\]
\[
\left\{\epsilon_{rs}\,{\cal G}^0_{t_1}(b,y_1)\,\sqrt h h^{rq}\overleftrightarrow{\frac{\partial}{\partial y_1^q}}\,{\cal G}^0_{t_2}(y_1,y_2)\,\sqrt h h^{sp}\overleftrightarrow{\frac{\partial}{\partial y_2^p}}\,{\cal G}^0_{T-t_1-t_2}(y_2,a)\,
\right\}\Big|_{y_1=y_2=y}\,.
\]
$Z$ is fixed by $\langle \,1\,\rangle_{C}=1$ to be $1/A$.
The $\epsilon$ cut-off in the $t_2$ integration regulates the expression when the two insertions approach each other.
The insertions split the interval $(0,T)$ into three, at least one of which must have a length of the order of $T$.
Since for large $ t$ the heat-kernel ${\cal G}^0_t\sim 1/A$ which is independent of position the integrand will vanish when 
two adjacent intervals are of the order of $T$ so the integral only receives contributions when both insertions are close to end-points or when both inertions are close to each other but far from either end-point. Consequently as $T\rightarrow\infty$
\[
\int_0^T dt_1\,\int_\epsilon^T dt_2\,\epsilon_{rs}\,{\cal G}^0_{t_1}(b,y_1)\,\sqrt h h^{rq}\overleftrightarrow{\frac{\partial}{\partial y_1^q}}\,{\cal G}^0_{t_2}(y_1,y_2)\,\sqrt h h^{sp}\overleftrightarrow{\frac{\partial}{\partial y_2^p}}\,{\cal G}^0_{T-t_1-t_2}(y_2,a)\,
\]
\[
=\int_0^\infty dt_1\,\int_0^\infty dt_2\,
\,\epsilon_{rs}\,{\cal G}^0_{t_1}(b,y_1)\,\sqrt h h^{rq}\overleftarrow{\frac{\partial}{\partial y_1^q}}\,\frac{1}{A}\,\sqrt h h^{sp}{\frac{\partial}{\partial y_2^p}}\,{\cal G}^0_{t_2}(y_2,a)\,
\]
\[
+\int_0^\infty dt_1\,\int_\epsilon^\infty dt_2\,
\epsilon_{rs}\,{\cal G}^0_{t_1}(b,y_1)\,\sqrt h h^{rq}\overleftrightarrow{\frac{\partial}{\partial y_1^q}}\,{\cal G}^0_{t_2}(y_1,y_2)\,\sqrt h h^{sp}\overleftarrow{\frac{\partial}{\partial y_2^p}}\,\frac{1}{A}\,
\]
\[
+\int_0^\infty dt_1\,\int_\epsilon^\infty dt_2\,\epsilon_{rs}\,\frac{1}{A}\,\sqrt h h^{rq}{\frac{\partial}{\partial y_1^q}}\,{\cal G}^0_{t_2}(y_1,y_2)\,\sqrt h h^{sp}\overleftrightarrow{\frac{\partial}{\partial y_2^p}}\,{\cal G}^0_{t_1}(y_2,a)
\]
\[
+\int_0^\infty dt_1\,\int_\epsilon^\infty dt_2\,
\epsilon_{rs}\,\frac{1}{A}\,\sqrt h h^{rq}{\frac{\partial}{\partial y_1^q}}\,{\cal G}^0_{t_2}(y_1,y_2)\,\sqrt h h^{sp}\overleftarrow{\frac{\partial}{\partial y_2^p}}\,\frac{1}{A}\,.
\]
As we need to set $y_1=y_2$ the last integrand is seen to vanish by using the spectral decomposition of the heat kernel because
\[
\epsilon^{qp}{\frac{\partial}{\partial y_1^q}}\,{\cal G}^0_{t_2}(y_1,y_2)\,\overleftarrow{\frac{\partial}{\partial y_2^p}}\Big|_{y_1=y_2=y}=\sum_\lambda e^{-\lambda t_2} \epsilon^{qp}\frac{\partial u_\lambda}{\partial y^q}\frac{\partial u_\lambda}{\partial y^p}=0\,.
\]
Using (\ref{avdeltaint}) and (\ref{solvpsi}) this gives
\[
\langle \,n[C]\,\rangle_{C}=
\]
\[
-i\int \frac{d^2y}{16\pi^2}\Big\{\left(\frac{1}{\bar y-\bar b}+\frac{1}{y-\bar b}\right)\left(\frac{1}{y-a}+\frac{1}{\bar y-a}\right)
-\left(\frac{1}{\bar y-\bar a}+\frac{1}{y-\bar a}\right)\left(\frac{1}{y-b}+\frac{1}{\bar y-b}\right)\]
\[
+\frac{1}{y-\bar y}\left(\frac{1}{\bar y-\bar b}+\frac{1}{y-\bar b}+\frac{1}{y-b}+\frac{1}{\bar y-b}
-\frac{1}{\bar y-\bar a}-\frac{1}{y-\bar a}-\frac{1}{y-a}-\frac{1}{\bar y-a}\right)\Big\}\,.
\]
In deriving this we have used the result that terms of the form $1/(y_1-y_2)$ that would diverge when $y_1=y_2=y$ actually vanish when regulated by $\epsilon$ essentially because this regularisation replaces the singular terms in 
$\int {\cal G}_t^0(y_1-y_2) dt $ by a power series in integer powers of $(y_1-y_2)^a(y_1-y_2)^b\delta_{ab}/\epsilon$ whose derivative with respect to $y_1$ vanishes at $y_1=y_2$.

\section{Supersymmetric generalisation}

It was shown in the Section 3 that the average of the number of intersections of two bosonic curves on a curved surface provided a way of continuing the path ordering of a field theory from the boundary into the bulk of the surface upon which the field lives. This is the first step to generalising the worldsheet model of \cite{Edwards:2014cga} to include non-abelian gauge theories. Of course the next step would be to include the self interactions of the gauge field which would have arisen from singularities in our intersection number model.  
We also still have the problem of divergences in the bulk arising from the coincidence of vertex operator insertions. It was shown that when supersymmetry is included on the worldsheet there exists sufficient structure to eliminate these divergences. Supersymmetry on the worldsheet also naturally includes a way of coupling of fermions to the gauge fields. It is therefore natural to generalise the non-abelian model to include supersymmetry.

One may then consider the  supersymmetric analogue for the intersection of two curves by replacing the surface by a two dimensional supermanifold with coordinates $(z,\bar{z},\theta,\bar{\theta})$, with $\theta$ Grassmann odd and related to $z$ by a supersymmetry transformation. If the end points of the two curves are now $(b_i,\theta_i(b))$ and $(a_i,\theta_i(a))$ with $i=1,2$, then we might expect (\ref{boundprop}) to become a relation of the form 
\[
\langle n_F[C_1,\,C_2]\rangle_{C_1,C_2}=\,\frac{(b_1-b_2-\theta_1(b)\theta_2(b))}{|b_1-b_2-\theta_1(b)\theta_2(b)|}
\]
which would result from an intersection number of the form
\[
n_F[C_1,C_2]=-i\int_{C1,C2}\left(dz_1d\bar z_2-d\bar z_1dz_2\right)
 \delta_c(z_1-z_2-\theta_1(b)\theta_2(b))\,.
 \label{n gauge}
\]
This is a straightforward generalisation of the number of intersections of bosonic curves with a shift of the displacement $z_1-z_2$ by the constant $-\theta_1(b)\theta_2(b)$. We will show in this section how this arises as the gauge fixed form of a more general intersection number with $\theta_1$ and $\theta_2$ dynamical by considering the intersection of two curves on a compact curved supermanifold that specularly reflect at the boundary. We will discuss an appropriate action functional for a spinning particle coupled to background supergravity that will be used to weight each curve upon path integration. This action has sufficient symmetries to allow a gauge fixing procedure in which the gauge fixed weight takes the same form as the bosonic weight used previously so that functional integrals of gauge invariant 
quantities reduce to bosonic functional integrals that have already been evaluated. We must therefore make sure that the number of intersections shares this gauge symmetry and so we make a slight modification to the form of $n_F$ to ensure this.

\subsection{The Green-Schwarz superparticle}

In the bosonic case the long random curves were naturally interpreted as the worldlines of bosonic point particles in the $T\rightarrow \infty$ limit. Here we will interpret the curves as the worldlines of superparticles on a curved supermanifold. Average quantities are then calculated by summing over curves weighted by the action of a superparticle coupled to 2d supergravity. This action is required to share the supersymmetry of the underlying supermanifold as the target space of the superparticle is the worldsheet of the spinning string. An action that satisfies this criterion is that of the   Green-Schwarz (GS) superparticle. The GS superparticle has manifest target space supersymmetry and, in 2 dimensions, has sufficient symmetry to allow us to remove most of the $\theta$ dependence from the action all together. The form of the Lagrangian we use is similar to that studied in \cite{Knutt} 
except that we are working on a Euclidean surface and have introduced an extra mass parameter, $\mu$. The Lagrangian of the GS superparticle in flat 2 dimensional superspace is then
\[
L_0= \ \frac{\pi\bar{\pi}}{\sqrt{g}}-\mu(\theta\dot{\bar{\theta}}+\bar{\theta}\dot{\theta})+\mu^2\sqrt{g} \label{Lag1}
\]
where $\pi\equiv \dot{z}+\theta\dot{\theta}$ and $\bar{\pi}\equiv \dot{\bar{z}}+\bar{\theta}\dot{\bar{\theta}}$ are the globally supersymmetric generalisations of $\dot{z}$ and $\dot{\bar{z}}$ respectively. The Lagrangian is invariant under reparametrisations, $t\rightarrow \tau(t)$ and the global supersymmetric variations: $\delta z=-\eta \theta, \ \delta \theta=\eta$. The action also possesses an additional worldline kappa symmetry of the form
\[
\delta_{\kappa}z=-\theta\delta_{\kappa}\theta, \ \ \ \delta_{\kappa}\theta=-\left(\kappa+\frac{\bar{\kappa}\pi}{\mu\sqrt{g}}\right), \ \ \ \delta_{\kappa}\bar{z}=-\bar{\theta}\delta_{\kappa}\bar{\theta}, \ \ \ \delta_{\kappa}\bar{\theta}=-\left(\bar{\kappa}+\frac{\kappa\bar{\pi}}{\mu\sqrt{g}}\right) \,,\label{kappa 1}
\]
\[
\delta_{\kappa}\sqrt{g}=\frac{2}{\mu}(\dot{\bar{\theta}}\kappa+\dot{\theta}\bar{\kappa})\,.
\]
We can use this symmetry to choose a gauge in which $\dot{\theta}=\dot{\bar{\theta}}=0$. We can also use the reparametrisation invariance to fix $\sqrt{g}=T$ as in the bosonic case. In this gauge (\ref{Lag1}) then reduces to the Lagrangian of the massive free bosonic particle. In the bosonic model we used the action for a free massless particle and so setting $\mu=0$ reduces the Lagrangian to (\ref{action}). We are interested in coupling the superparticle to supergravity as we would like to generalise the model of \cite{Edwards:2014cga} to curved space. The superparticle coupled to supergravity is best described in terms of forms and so we first write (\ref{Lag1}) as
\[
L_0=\frac{\dot{e}^z\dot{e}^{\bar{z}}}{\sqrt{g}}+2\mu \dot{e}^A\Gamma_A+\mu^2\sqrt{g} \label{Lag 2}
\]
where $\dot{e}^A=\dot{z}^Me_M^{ \ A}$, $e_M^{ \ A}$ is the flat super-vielbein and $\Gamma_A$ are gauge fields. In particular we have $\dot{e}^z=\pi$, $\dot{e}^{\bar{z}}=\bar{\pi}$, $\dot{e}^{\theta}=\dot{\theta}$, $\dot{e}^{\bar{\theta}}=\dot{\bar{\theta}}$, $\Gamma_{\theta}=\bar{\theta}/2$ and $\Gamma_{\bar{\theta}}=\theta/2$. With the Lagrangian in this form it is easy to generalise to curved space by promoting the flat space super-vielbein, $e$, to the curved super-vielbein, $E$, so that the Lagrangian of the superparticle coupled to supergravity is
\[
L_1=\frac{\dot{E}^z\dot{E}^{\bar{z}}}{\sqrt{g}}+2\mu\dot{E}^A\Gamma_A+\mu^2\sqrt{g} \label{Lag 3}
\]
where again $\dot{E}^A=\dot{z}^ME_M^{ \ A}$.
The supergravity covariant derivative is
\[
\nabla_A=E_A^MD_M+\Omega_A \label{cov der}
\]
where $D_A=(D_{\theta},D_{\bar{\theta}})$ are the flat superderivatives and $\Omega_A=\omega_AM$ is the spin connection.
The $\Gamma_{A}$ are subject to the supergravity constraints
\[
\nabla_{\theta}\Gamma_{\bar{\theta}}+\nabla_{\bar{\theta}}\Gamma_{\theta}=1, \ \ \ \ \Gamma_z=\nabla_{\theta}\Gamma_{\theta}, \  \ \ \ \Gamma_{\bar{z}}=\nabla_{\bar{\theta}}\Gamma_{\bar{\theta}}
\]
and the covariant derivatives satisfy
\[
\nabla_z=\frac{1}{2}\{\nabla_{\theta},\nabla_{\theta}\} \ \ \ \nabla_{\bar{z}}=\frac{1}{2}\{\nabla_{\bar{\theta}},\nabla_{\bar{\theta}}\}\,.
\]
These constraints are solved in superconformal gauge using a compensator function, $S$, so that
\[
\nabla_{\theta}=e^S[D+2(DS)M] \ \ \ \nabla_{\bar{\theta}}=e^S[D-2(\bar{D}S)M]
\]
\[
\nabla_z=e^S[\partial+2(DS)D+2(\partial S)M]
\]
\[
\nabla_{\bar{z}}=e^S[\bar{\partial}+2(\bar{D}S)\bar{D}-2(\bar{\partial} S)M]\,.
\]
Switching to the co-ordinate basis in which the covariant derivative is $\nabla_A=\mathcal{E}_A^{ \ M}\partial_M+\Omega_A$ allows us to read off the elements of the inverse supervielbein and invert to obtain
\[ \mathcal{E}_M^{ \ A}=\left( \begin{array}{cccc}
e^{-2S} & 0 & -2e^{-S}DS & 0 \\
0 & e^{-2S} & 0 & -2e^{-S}\bar{D}S \\
-e^{-2S}\theta & 0 & e^{-S}[1-2(DS)\theta] & 0 \\
0 & -e^{-2S}\bar{\theta} & 0 & e^{-S}[1-2(\bar{D}S)\bar{\theta}] \end{array} \right)\] 
and compute $\text{sdet}(\mathcal{E})=e^{-2S}$. 
One can then write (\ref{Lag 3}) in terms of superspace co-ordinates as
\[
L_1=e^{-4S}\frac{\pi\bar{\pi}}{\sqrt{g}}+2\mu e^{-S}\bigg(\pi(D\Gamma_{\theta}-(DS)\Gamma_{\theta})+
\bar{\pi}(\bar{D}\Gamma_{\bar{\theta}}-(\bar{D}S)\Gamma_{\bar{\theta}})\bigg) \nonumber
\]
\[
- 2\mu e^{-S}(\Gamma_{\theta}\dot{\theta}+\Gamma_{\bar{\theta}}\dot{\bar{\theta}})+\mu^2\sqrt{g}\label{full}\,.
\]
This can be simplified by introducing $G\equiv e^{-S}\Gamma_{\theta}$ and $\bar{G}\equiv e^{-S}\Gamma_{\bar{\theta}}$   allowing us to write the Lagrangian as
\[
L_1=e^{-4S}\frac{\pi\bar{\pi}}{\sqrt{g}}+2\mu(\pi DG+\bar{\pi}\bar{D}\bar{G})-2\mu(G\dot{\theta}+\bar{G}\dot{\bar{\theta}})+\mu^2\sqrt{g} \,.\label{Lag 4}
\]
The supergravity constraint on $\Gamma_{\theta}$ and $\Gamma_{\bar{\theta}}$ now becomes a constraint on $G$ and $\bar{G}$ and takes the form
\[
D\bar{G}+\bar{D}G=e^{-2S} \,.\label{constraint}
\] 
The change of the Lagrangian under a general variation of the form $\delta z=-\theta\delta \theta$ is
\[
\delta L=2e^{-4S}\frac{\pi\bar{\pi}}{\sqrt{g}} \ \big(\delta\bar{\theta}\bar{D}(-2S)+\delta\theta D(-2S)\big)-\frac{2e^{-4S}}{\sqrt{g}} \ \big(\pi\dot{\bar{\theta}}\delta\bar{\theta}+\bar{\pi}\dot{\theta}\delta\theta\big) \nonumber
\]
\[
 -2\mu e^{-2S}(\delta\bar{\theta}\dot{\theta}+\delta\theta\dot{\bar{\theta}})-2\mu e^{-2S} \ \big(\pi\delta\bar{\theta}D(-2S)+\bar{\pi}\delta\theta\bar{D}(-2S)\big) \nonumber
\]
\[
-\delta\sqrt{g}\bigg(e^{-4S}\frac{\pi\bar{\pi}}{g}-\mu^2\bigg)\,.\nonumber
\]
\newpage
The Lagrangian is then kappa invariant if
\[
\delta_{\kappa} \theta=-\bigg(\kappa e^{2S}+\frac{\bar{\kappa}\pi}{\mu\sqrt{g}}\bigg), \ \ \ \ \delta_{\kappa}\bar{\theta}=-\bigg(\frac{\kappa\bar{\pi}}{\mu\sqrt{g}}+\bar{\kappa}e^{2S}\bigg)\,, \nonumber
\]
\[
\delta_{\kappa}\sqrt{g}=\frac{2\dot{\bar{\theta}}\kappa}{\mu}-\frac{2\kappa\bar{\pi}\bar{D}(-2S)}{\mu}+\frac{2\dot{\theta}\bar{\kappa}}{\mu}-\frac{2\bar{\kappa}\pi D(-2S)}{\mu}\,.\nonumber
\]

As in the bosonic model we wish to consider the massless limit of the superparticle. Taking $\mu=0$ gives the Lagrangian
\[
L_2=e^{-4S}\frac{\pi\bar{\pi}}{\sqrt{g}}\,.
\]
The massless kappa transformations are obtained by setting $\kappa'\equiv \kappa/\mu$ and then letting $\mu\rightarrow 0$.
\[
\delta_\kappa \theta=-\frac{\bar{\kappa}'\pi}{\sqrt{g}}, \ \ \ \delta_\kappa \bar{\theta}=-\frac{\kappa'\bar{\pi}}{\sqrt{g}}, \ \ \ \delta_\kappa z=-\theta\delta_\kappa\theta, \ \ \ \delta_\kappa\bar{z}=-\bar{\theta}\delta_\kappa\bar{\theta} \nonumber
\]
\[
\delta_{\kappa}\sqrt{g}=2\dot{\bar{\theta}}\kappa'-2\kappa'\bar{\pi}\bar{D}(-2S)+2\dot{\theta}\bar{\kappa}'-2\bar{\kappa}'\pi D(-2S)\,.\nonumber
\]
One can now use the kappa transformations to pick a gauge in which $\dot{\theta}=\dot{\bar{\theta}}=0$ by requiring that $\theta(t)=\theta(b)$ and $\bar{\theta}(t)=\bar{\theta}(b)$ for all $t$. Gauge fixing $\theta_i$ and denoting the gauge fixed form of  $-4S(z,\theta(b))$ as $\tilde{\phi}(z)$ we find that the Lagrangian reduces to
\[
L_2'=\frac{1}{\sqrt{g}}e^{\tilde{\phi}}\dot{z}\dot{\bar{z}}
\]
which is the Lagrangian of the free massless bosonic particle in conformal gauge. To this we should add Faddeev-Popov terms associated with the fixing of the reparametrisation invariance and kappa symmetry
\[
 \bar\lambda\left(\theta(b)-\theta\right)+\lambda\left(\bar\theta(b)-\bar\theta\right)
 +\bar B\frac{C\bar{\pi}}{\sqrt{g}}+B\frac{\bar C{\pi}}{\sqrt{g}}\,.
\]
$\lambda$ acts is a Lagrangian multiplier imposing the gauge condition and the ghosts $B$ and $C$ generate the Faddeev-Popov determinant of a local quantity (as opposed to a differential operator) 
which can be ignored. 

The observables that we work with should be BRST invariant. For example in the bosonic case we focus on long curves by inserting $\delta(\int^1_0\sqrt g \,d\xi -T)$ into the functional integral. This is reparametrisation invariant, which is sufficient in the bosonic case, but it is not $\kappa$ invariant which we also need in the supersymmetric case. However the kappa variation of 
\[
 \sqrt g\left( 1-(\bar\theta-\bar\theta(b))\bar D(-2S)-(\theta-\theta(b)) D(-2S)\right)\equiv \zeta
\]
is zero when we impose the gauge conditions so we can use this to make a BRST invariant insertion.

\subsection{Supersymmetric intersection number}
Now that we have an appropriate action with which we can weight each curve by, we must now consider a supersymmetric generalisation of the intersection number.
 The underlying curved supermanifold should be be interpreted as the worldsheet of the spinning string. Quantities on the supermanifold should then share the supersymmetry of the supermanifold itself, as the action in flat space does
The number of intersections on this surface should also have this supersymmetry. We will introduce supersymmetry into our model by first rewriting the number of intersections of bosonic curves in terms of a quantity that shares the symmetries of the surface upon which it is defined. For the bosonic case we have rotational and translational symmetry and so the quantity of interest is the displacement, $s\equiv z_1-z_2$. Introducing $\xi_1$ and $\xi_2$ to parametrise the curves $C_1$ and $C_2$ respectively, the number of intersections of two curves can then be written as
\[
n[C_1,C_2]=i\int_{C_1,C_2}  \ \delta_c^2(s)(\dot{s}\bar{s}'-\dot{\bar{s}}s') \  \ d\xi_1 d\xi_2
\]
where $\dot{s}\equiv dl/d\xi_1 $ and $s'\equiv dl/d\xi_2 $. On the supermanifold we have the additional Grassmann odd co-ordinates, $\theta$, related to the 'bosonic' co-ordinates by the supersymmetry transformation $\delta z=-\eta\theta$ and $\delta \theta=\eta$. The natural generalisation of $s$ that is invariant under these transformations is $l\equiv z_1-z_2-\theta_1\theta_2$. 
A first step to generalising the intersection number would then be to replace $s$ with $l$, so that the number of intersections of two fermionic curves would be
\[
n_F[C_1,C_2]=i\int_{C_1,C_2} \delta_c^2(l)(\dot{l}\bar{l}'-\dot{\bar{l}}l') \ d\xi_1d\xi_2 \label{int number}
\]
Averaging this functional over the two curves requires summing over curves weighted by the action (\ref{full}). A nice feature of the Green-Schwarz action was the existence of the worldline $\kappa$ symmetry which allowed us to choose a gauge in which the action reduced to that of the massless free bosonic particle. Unfortunately, as $l$ is not $\kappa$ invariant, the number of intersections is also not $\kappa$ invariant, meaning we are unable to carry out the gauge fixing on the functional integral which would have vastly simplified the calculation. We can however modify the displacement to make it $\kappa$ invariant by defining $L\equiv z_1-z_2+\theta_1(b)\theta_1-\theta_2(b)\theta_2-\theta_1(b)\theta_2(b)$. $L$ has the property of invariance once the gauge conditions are imposed. We then propose that the supersymmetric and $\kappa$ invariant number of intersections is
\[
n_F[C_1,C_2]=i\int_{C_1,C_2} \ (\pi_1^0\bar{\pi}^0_2-\bar{\pi}_1^0\pi_2^0) \ \delta^2(L) \ d\xi_1d\xi_2
\]
with $\pi_i^0\equiv \dot{z}_i+\theta_i(b)\dot{\theta}_i$. We will consider averaging this functional over both curves and keeping all four end points fixed to begin with. We will then consider averaging over one  of the end points of each curve using a gauge fixed volume element. 
With the end points fixed we can then average any supersymmetric and $\kappa$ invariant functional, $\Omega[C_1]$, over $C_1$ by computing the functional integral
\[
\braket{\Omega}_{C_1}\equiv\frac{1}{Z}\int\mathcal{D}g\mathcal{D}z_1\mathcal{D}\theta_1 \mathcal{D} \lambda
\mathcal{D} B \mathcal{D} C
\ \delta\big(\int^1_0
\sqrt{\zeta }d\xi-T\big) \ \Omega[C_1] \ e^{-S_{FP}[g,z_1,\theta_1,\lambda,B,C]}\,,
\]
where $S_{FP}$ is the gauge fixing action.
The gauge conditions reduce the functional integral to
\[
\braket{\Omega}_{C_1}=\frac{1}{Z}\int\mathcal{D}z_1 \  \ \Omega'[C_1] \ e^{-S'[z_1]}\label{func int}
\]
with 
\[
S'[z]=\frac{1}{2}\int^T_0 dt \ e^{\tilde{\phi}}\dot{z}\dot{\bar{z}}
\]
and $\Omega'$ is the gauge fixed form of $\Omega$. (\ref{func int}) is then equivalent to the bosonic functional integral considered before. The average of the number of intersections is then
\[
\braket{n_F[C_1,C_2]}_{C_1}= \nonumber
\]
\[
-i\int_{C_2}\left\langle\int_{C_1}  \ \delta^2(z_1-z_2-\theta_1(b)\theta_2(b))dz_1 \right\rangle_{C_1} d\bar{z}_2 \nonumber
\]
\[
+i\int_{C_2}\left\langle\int_{C_1} \ \delta^2(z_1-z_2-\theta_1(b)\theta_2(b))d\bar{z}_1\right\rangle_{C_1} dz_2\,. \label{n_F av}
\]
The averages can be done by defining the new variable, $z_2'$ as a shift of $z_2$ such that $z_2'=z_2+\theta_1(b)\theta_2(b)$. 
Reverting to general co-ordinates $x_1$ and $x_2'$ to make contact with the bosonic calculation and introducing a Fourier decomposition of the delta function, we have for the first average
\[
\left\langle \int_{C_1} \delta^2(x_1-x_2')dx_1^{u}\right\rangle_{C_1}=\nonumber
\]
\[
\left\{\frac{\delta}{\delta {\cal A}_u({x_2'})}\frac{1}{Z}
\int {\cal D} {x_1}\,\,e^{-\int_0^T \left(\frac{1}{2}h_{rs}(x_1)\,\dot x^r_1\,\dot x_1^s   -{\cal A}_r\dot x_1^r\right)\,dt
}
\right\}\Big |_{{\bf \cal A}=0}\,.
\]
By analogy with the bosonic case this is
\[
=\frac{1}{2} \int_0^{\infty} dt\, \big(-\sqrt{h}h^{rs}\partial_{s}\mathcal{G}^{'  0}_{t}(b_1,x_2')+\sqrt{h}h^{rs}\partial_{s}\mathcal{G}^{'  0}_{T-t}(x_2',a_1))\big)
\label{res}
\]
after fixing the normalisation constant and taking the $T\rightarrow \infty$ limit. Now $\mathcal{G}^{'  0}_{t}(x_1,x_2')$ is not equivalent to the bosonic heat kernel used in the previous section due to subtleties involving the $\theta$ co-ordinates. Here we have 
\[
{\cal G}^{'0}_T(x_1,x_2') =\langle\,x_1,|\,e^{-T\hat H_0}\,|\,x_2'\,\rangle
+\langle\,x_1\,|\,e^{-T\hat H_0}\,|\,x_2'^R\,\rangle\,.
\]
We are still considering specular reflections of the curves when they reach the boundary but in this case the reflected coordinate of the i'th curve is $(x_i^R,\theta_i^R)$ and so $x_2'^R=x_2^R+\theta_1(b)\theta_2^{R}(b)$. 
The integral of ${\cal G}^{'0}_T$ over $t$ results in a generalisation of the Green function discussed earlier, denoted $G'$:
\[
\int_0^\infty dt \ \bigg({\cal G}'_T(z_1,z_2')-\frac{1}{A}\bigg)=2G'(z_1,z_2')\,.
\]
It satisfies
\[
-4e^{-\tilde{\phi}}\bar{\partial}\partial G'=-2ie^{-\tilde{\phi}}\delta^2(z_1-z_2-\theta_1(b)\theta_2(b))-\frac{1}{A} \label{susy green}
\]
and modified Neumann conditions
\[
(\partial_i-\bar{\partial}_i)G'(z_i,z_j,\theta_i,\theta_j)|_{z_i=\bar{z}_i,\theta_i=\bar{\theta}_i}=0 \,.\label{susy Neumann}
\]
The solution to (\ref{susy green}) satisfying (\ref{susy Neumann}) is then
\[
G'=-\frac{1}{2\pi}\text{log}(|z_1-z_2-\theta_1(b)\theta_2(b)|)-\frac{1}{2\pi}\text{log}(|z_1-\bar{z}_2-\theta_1(b)\bar{\theta}_2(b)|)-\Psi(z_1,z_2,\theta_1(b),\theta_2(b)) \label{green def}
\]
where $\Psi$ solves
\[
-4e^{\tilde{\phi}(z_1,\theta_1(b))}\bar{\partial}_1\partial_1\Psi=-4e^{\tilde{\phi}(z_2,\theta_2(b))}\bar{\partial}_2\partial_2\Psi=\frac{1}{A} \label{psi}
\]
and the modified Neumann conditions.
Using these results we find 
\[
\left\langle\int_{C_1}  \ \delta^2(z_1-z_2') dz_1\right\rangle_{C_1}=\frac{\partial G'(b_1,z_2)}{\partial \bar{z}_2}-\frac{\partial G'(z_2,a_1)}{\partial \bar{z}_2}\,. \label{fixed av}
\]
From (\ref{psi}) it is clear that $\Psi$ can be decomposed as
$\Psi(z_1,z_2,\theta_1,\theta_2)=f(z_1,\theta_1)+f(z_2,\theta_2)$. Because of this, (\ref{fixed av}) is independent of the zero mode contribution to the Green function. Using (\ref{green def}) we find
\[
\left\langle\int_{C_1}  \ \delta^2(z_1-z_2-\theta_1\theta_2(b))  \ dz_1\right\rangle_{C_1}= \nonumber
\]
\[
\frac{1}{4\pi}\bigg(\frac{1}{\bar{b}_1-\bar{z}_2-\bar{\theta}_1(b)\bar{\theta}_2(b)}+\frac{1}{b_1-\bar{z}_2-\theta_1(b)\bar{\theta}_2(b)}\bigg) \nonumber
\]
\[
-\frac{1}{4\pi}\bigg(\frac{1}{\bar{a}_1-\bar{z}_2-\bar{\theta}_1(b)\bar{\theta}_2(b)}+\frac{1}{a_1-\bar{z}_2-\theta_1(b)\bar{\theta}_2(b)} \bigg)\,.
\]
The number of intersections of two curves, $C_1$ and $C_2$, on a supermanifold, averaged over $C_1$ is then
\[
\braket{n_F[C_1,C_2]}_{C_1}=-\frac{1}{2\pi}\Im \text{log}_{2^*}\bigg(\frac{(b_1-b_2-\theta_1(b)\theta_2(b))(\bar{b}_1-b_2-\bar{\theta}_1(b)\theta_2(b))}{(b_1-a_2-\theta_1(b)\theta_2(b))(\bar{b}_1-a_2-\bar{\theta}_1(b)\theta_2(b))}\bigg) \nonumber
\]
\[
+\frac{1}{2\pi}\Im \text{log}_{2^*}\bigg(\frac{(a_1-b_2-\theta_1(b)\theta_2(b))(\bar{a}_1-b_2-\bar{\theta}_1(b)\theta_2(b))}{(a_1-a_2-\theta_1(b)\theta_2(b))(\bar{a}_1-a_2-\bar{\theta}_1(b)\theta_2(b))}\bigg)-\big(n[C_2,C_1^*]-n[C_2^*,C_1^*]\big)\,.\nonumber
\]
The number of intersections of $C_2^*$ and $C_1^*$ is a straight forward generalisation of the bosonic case and can be obtained by  shifting $x_2$ as before, we have
\[
n[C_2^*,C_1^*]= \frac{1}{2\pi}\Im \text{log}_{2*}\bigg(\frac{(b_1-b_2-\theta_1(b)\theta_2(b))(a-a_2-\theta_1(b)\theta_2(b))}{(b_1-a_2-\theta_1(b)\theta_2(b))(a-b_2-\bar{\theta}_1(b)\theta_2(b))}\bigg)\nonumber
\]
\[
-\frac{1}{2\pi}\Im \text{log}_{1*}\bigg(\frac{(b_1-b_2-\theta_1(b)\theta_2(b))(a-a_2-\theta_1(b)\theta_2(b))}{(b_1-a_2-\theta_1(b)\theta_2(b))(a-b_2-\theta_1(b)\theta_2(b))}\bigg)\,.
\]
\newpage
We also have
\[
\braket{n_F[C_2,C_1^*]}_{C_2}=-\frac{1}{2\pi}\Im \text{log}_{1*}\bigg(\frac{(b_1-b_2-\theta_1(b)\theta_2(b))(b_1-\bar{b}_2-\theta_1(b)\bar{\theta}_2(b))}{(a-b_2-\theta_1(b)\theta_2(b))(a-\bar{b}_2-\theta_1(b)\bar{\theta}_2(b))}\bigg) \nonumber
\]
\[
+\frac{1}{2\pi}\Im \text{log}_{1*}\bigg(\frac{(b_1-a_2-\theta_1(b)\theta_2(b))(b_1-\bar{a}_2-\theta_1(b)\bar{\theta}_2(b))}{(a-a_2-\theta_1(b)\theta_2(b))(a-\bar{a}_2-\theta_1(b)\bar{\theta}_2(b))}\bigg)\,. \nonumber
\]
Using these results we find 
\[
\left\langle\left\langle n_F[C_1,C_2]\right\rangle_{C_1}\right\rangle_{C_2}=\frac{1}{2\pi}\Im \text{log}_{1*}\bigg(\frac{(b_1-\bar{b}_2-\theta_1(b)\bar{\theta}_2(b))(a_1-\bar{a}_2-\theta_1(b)\bar{\theta}_2(b))}{(a_1-\bar{b}_2-\theta_1(b)\bar{\theta}_2(b))(b_1-\bar{a}_2-\theta_1(b)\bar{\theta}_2(b))}\bigg) \nonumber
\]
\[
-
\frac{1}{2\pi}\Im \text{log}_{2*}\bigg(\frac{(\bar{b}_1-b_2-\bar{\theta}_1(b)\theta_2(b))(\bar{a}_1-a_2-\bar{\theta}_1(b)\theta_2(b))}{(\bar{b}_1-a_2-\bar{\theta}_1(b)\theta_2(b))(\bar{a}_1-b_2-\bar{\theta}_1(b)\theta_2(b))}\bigg)\,.
\]

At this point we can consider integrating over one of the end points. To do this we need the reduced volume element of just the bosonic co-ordinates. We can obtain this by writing down the line element and imposing the gauge conditions
\[
ds^2=\eta_{AB}\mathcal{E}_{M}^{ \ A} dz^M \mathcal{E}_{N}^{ \ B}dz^N=\mathcal{E}_{z}^{ \ \bar{z}}\mathcal{E}_{\bar{z}}^{ \ z} dz d\bar{z}+\frac{1}{2}\{\mathcal{E}_{z}^{ \ \theta},\mathcal{E}_{\bar{z}}^{ \ \bar{\theta}}\} dzd\bar{z} \nonumber
\]
\[
=e^{\tilde{\phi}}dzd\bar{z} \,,
\]
therefore the invariant volume element for the i'th end point is $\sqrt{h} \ d^2a_i=e^{\tilde{\phi}}d^2a_i$. We can now integrate over $a_1$ and $a_2$ and let $(b_1,\theta_1)$ and $(b_2,\theta_2)$ approach the boundary. Call the result of these actions $G_F(x_1,\theta_1;x_2,\theta_2)$ so that 
\[
G_F=-\frac{(x_1-x_2-\theta_1\theta_2)}{2|x_1-x_2-\theta_1\theta_2|}+\tilde{F}(x_1,x_2,\theta_1\theta_2)\,. \label{susy green'}
\]
Differentiating with respect to $(x_1,\theta_1)$ gives
\[
D_1G_F=-(\theta_1-\theta_2)\delta(x_1-x_2)+D_1\tilde{F}\,.
\]
Integrating along the boundary requires
\[
0=\int_{-\infty}^{+\infty} dx_1 \int d\theta_1 \ D_1G_F=-1+\int_{-\infty}^{+\infty} dx_1 \int d\theta_1 \ \theta_1\frac{\partial}{\partial x_1} \tilde{F}\,.
\]
The integral on the RHS is exactly the same as in the bosonic case after integrating out $\theta_1$ and so this does hold. (\ref{susy green'}) is then the Green function equation for $D=\partial/\partial\theta+\theta\partial/\partial x$ on a closed loop. This is a suitable supersymmetric generalisation of the bosonic case that one can use to introduce path ordering into the interior of the spinning string model.

\section{Conclusions}

We have studied certain contact interactions between world-lines on a curved surface motivated by earlier work on a string representation of the Wilson loop for Abelian gauge theory that is built on a string contact interaction for tensionless strings. We have found the average intersection number for two world-lines with fixed end-points and also the average intersection number for a single world-line with fixed end-points. The average was constructed using the natural world-line action of \cite{Brink:1976uf}. Taking the length of the world-line to infinity using the intrinsic metric in this action is an analogue of the tensionless limit of the string model because both remove the scale from the model. Consequently the average intersection numbers for curves with fixed end-points were found to be independent of the metric of the curved surface. When one end-point of each of two curves is integrated over their average intersection number is a function of the positions of the remaining fixed end-points and 
coincides with the propagator for a topological field theory constructed to provide a way of extending the notation of path-ordering around a closed curve onto a surface bounded by that curve. This is potentially of value in extending the string representation of the Wilson-loop to non-Abelian gauge theories as the string contact interaction takes place on the body of the world-sheet and but has to give rise to a path-ordered expression on the boundary. We have also discussed the supersymmetric generalisation as world-sheet supersymmetry is a necessary ingredient in the string model and found a natural generalisation of the bosonic result.

\acknowledgments
PM is grateful to STFC for support under the rolling grant 
ST/L000407/1 and CC is grateful to STFC for a studentship.
This research is also supported by the Marie Curie network GATIS 
(gatis.desy.eu) of the European Union's Seventh Framework Programme FP7/2007-2013/ under REA Grant Agreement No 317089.

\end{document}